\documentclass[noshowpacs,prl,aps,superscriptaddress,floatfix,twocolumn]{revtex4-2} 
\usepackage[utf8]{inputenc}
\usepackage{lineno,mathtools,url}
\usepackage{amsmath}
\usepackage{amssymb,mathrsfs}
\usepackage{physics}
\usepackage{gensymb}
\usepackage{graphicx}
\usepackage{hyperref}
\hypersetup{
     colorlinks=true,
     linkcolor=blue,
     filecolor=blue,
     citecolor = orange,      
   }
\usepackage[dvipsnames]{xcolor}
\usepackage{dsfont}
\usepackage{soul}
\usepackage{mathbbol} 
\usepackage[normalem]{ulem}
\allowdisplaybreaks
\def\sss{\scriptscriptstyle}

\begin{document}

\title{Topological Constraints on the Dynamics of Vortex Formation  in a Two-dimensional Quantum Fluid}

\author{T. Congy}
\affiliation{Department of Mathematics, Physics and Electrical Engineering, Northumbria University, Newcastle upon Tyne NE1 8ST, United Kingdom}
\author{P. Azam}
\author{R. Kaiser}
\affiliation{Institut de Physique de Nice, Université Côte d'Azur, CNRS, F-06560 Valbonne, France}
\author{N. Pavloff}
\affiliation{Universit\'e Paris-Saclay, CNRS, LPTMS, 91405, Orsay, France}
\affiliation{Institut Universitaire de France (IUF)}
\begin{abstract}
  We present experimental and theoretical results on formation of
  quantum vortices in a laser beam propagating in a nonlinear
  medium. Topological constrains richer than the mere conservation of
  vorticity impose an elaborate dynamical behavior to the formation
  and annihilation of vortex-antivortex pairs. We identify two such
  mechanisms, both described by the same fold-Hopf bifurcation. One of
  them is particularly efficient although it is not observed in the
  context of liquid helium films or stationary systems because it
  relies on the compressible nature of the fluid of light we consider
  and on the non-stationarity of its flow.
\end{abstract}

\maketitle

The propagation of light in a nonlinear medium can be described as a
(dispersive) hydrodynamic phenomenon. This approach, pioneered in the
1960's
\cite{Chiao1964,Talanov1965,Kelley1965,Talanov1966,Akhmanov1966,Akhmanov_1968}
and further developed in the 1990's
\cite{Coullet1989,Brambilla1991II,Pomeau1993,Staliunas1993} yielded
remarkable successes: observation of bright
\cite{Bjorkholm1974,Mollenauer1980,Barthelemy1985}, dark
\cite{Emplit1987,Krokel1988,Weiner1988,Conti2009}, cavity
\cite{Barland2002,Pedaci2006} and oblique \cite{Amo2011,Grosso2011}
solitons, of wave breaking and dispersive shock waves
\cite{Rothenberg1989,Wan2007,Xu2016,Bienaime2021,Azam2021}, of
quantized vortices
\cite{Arecchi1991,Swartzlander1992,Vaupel1996,Lagoudakis2008,Roumpos2011,Nardin2010,Nardin2011,Sanvitto2011,Anton2012,Dominici2015}
and of superfluid flow of light \cite{Amo2009,Michel2018}. An extreme
hydrodynamic-like behavior is the turbulent regime in which typical
observables display scale-invariant power law spectra in momentum
space. In the present study we focus on two dimensional (2D)
configurations, similar to those already studied in the field of
Bose-Einstein condensates, where quantum vortices proliferation but
also robust vortex structures have been observed
\cite{Neely2013,Gauthier2019,Johnstone2019,Reeves2022}. Although their
role in the different types of power laws which have been predicted
and/or observed \cite{Bradley2012,Neely2013,Galka2022} is not fully
elucidated \cite{Chesler2013,Nazarenko2014,Navon2016}, it makes no
doubt that understanding the dynamics of vortex formation is crucial
for unraveling the mechanisms leading to quantum turbulence. Recent
studies have demonstrated the efficiency of optical platforms for
studying this subject
\cite{Mamaev1996,Vocke2016,Maitre2021,Eloy2021,Sitnik2022,Panico2023,Abobaker2022}.

\begin{figure}
\centering
\includegraphics[width=\linewidth]{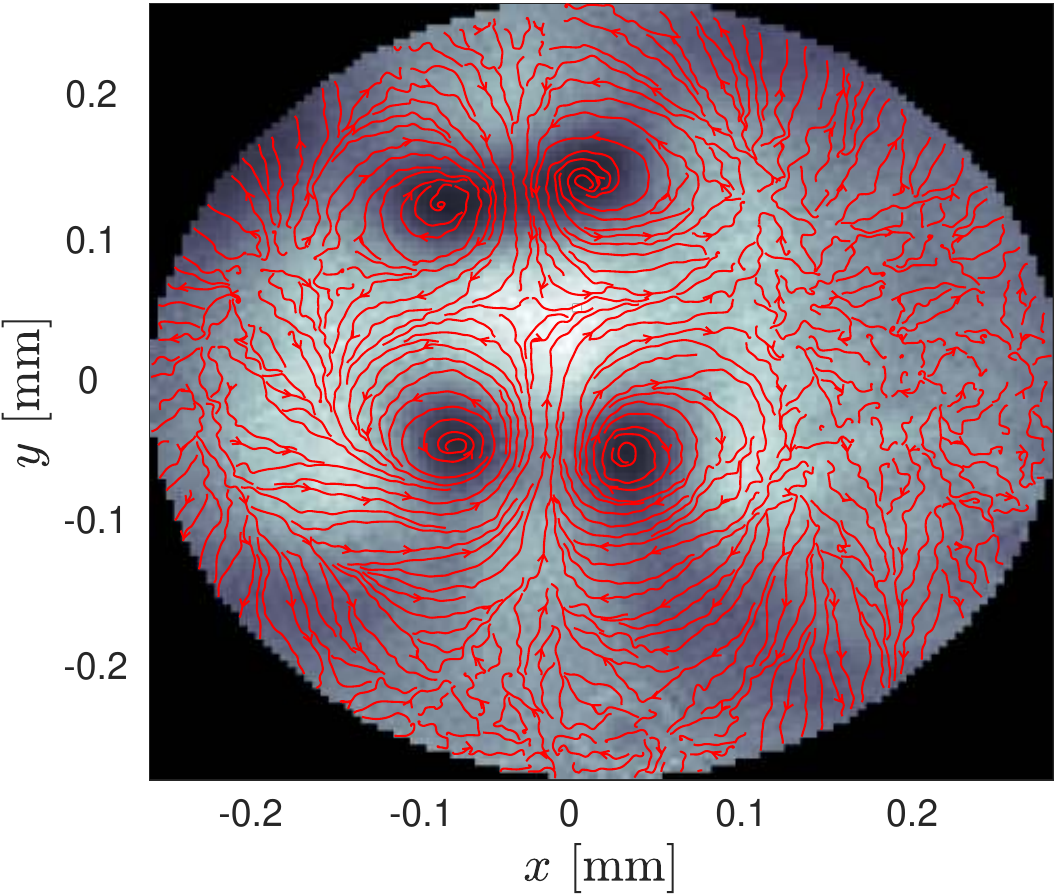}
\caption{Experimental intensity pattern and streamlines (in red) of the beam at the exit of the nonlinear vapor. Dark regions are of lesser intensity. One distinctly discerns two vortex-antivortex pairs and also a saddle located close to the origin.}\label{fig1}
\end{figure}

In the present work we use a nonlinear optical setup
\cite{Santic2018,Azam2021,Azam2022} for studying the formation and
annihilation of vortices and of other less conspicuous features, such
as saddles and phase extrema which also carry a topological
charge. Although the existence of these other critical points has a
long history \cite{Nye1974,Nye1981} their role in enforcing
topological constrains
\cite{Freund1995,Karman1997,Berry1998,Rozanov2004} is often
overlooked. Our detection tool is able to simultaneously record the intensity and the phase of a light sheet and then to reconstruct the streamlines of the flow of the fluid of light, as illustrated in Fig. \ref{fig1}.
This enables to investigate the formation mechanisms of vortices and
critical points. In particular we experimentally demonstrate for the
first time a scenario of vortex-antivortex formation first proposed
by Nye {\it et al.} in 1988 \cite{Nye1988} and identify a new one,
which appears simpler and presumably more efficient in the
time dependent flow of a compressible quantum fluid.

Consider a quantum fluid described by a scalar order parameter of the form
\begin{equation}
\label{eq:psi}
    \psi(\vec{r}\,)=A(\vec{r}\,)\exp[i S(\vec{r}\,)],
\end{equation}

defined in the plane ($\vec{r}=x\,\vec{e}_x +y \, \vec{e}_y$). In such
a system the formation of vortices is constrained by topological
rules: it is for instance well known that in the absence of externally
imparted angular momentum, vortices typically appear in pairs with
opposite quantized vorticity. This scenario is enriched by other
constrains \cite{Nye1988} originating from the fact that, to any
closed curve $C$ of the plane, are associated not one, but {\it two}
topological indices: the vorticity
$I_{\rm V}(C) =\frac{1}{2\pi} \oint_C {\rm d}S$ and the
Poincar\'e-Hopf index
$I_{\rm P}(C) =\frac{1}{2\pi} \oint_C {\rm d}\theta$, where $\theta$
is a polar angle of the ``velocity field''
$\vec{v}(\vec{r}\,)=\vec{\nabla} S$ and in both cases the integral is
performed clockwise. $I_{\rm V}$ and $I_{\rm P}$ are (positive or
negative) integers. This stems from the fact that along a close
contour the phase $S$ of the order parameter~\eqref{eq:psi} and the
orientation $\theta$ of the velocity must both vary by integer
multiples of $2\pi$ \cite{note_OF}.  If $S$ is regular and
well-defined in the interior of $C$, then $I_{\rm V}(C)=0$.  This
value does not change unless a vortex \footnote{Vortices are also
  named wavefront dislocations \cite{Nye1974} or amphidromic points (a
  terminology used in oceanography).} crosses $C$. To each vortex one
can associate a vorticity and a Poincar\'e-Hopf index by integrating
along a small circle around the vortex core. This yields
$I_{\rm P}=+ 1$ and typically \footnote{In many physical situations
  vortices with $|I_{\rm V}|$ larger than unity are unstable.}
$I_{\rm V}=\pm 1$ for each vortex.  Besides vortices, other points are
also associated with a finite Poincar\'e-Hopf index: those at which
the velocity of the flow cancels.  They are known as critical points
or equilibria.  For a potential flow such as ours, where the phase $S$
is the velocity potential, they are of two types: phase extrema (local
maxima or local minima) and phase saddles.  For an extremum
$I_{\rm P}=+ 1$ and for a saddle $I_{\rm P}=- 1$ \footnote{Higher
  values of $I_{\rm P}$ can be reached
  \cite{Scheuermann1997,Freund2000}, but the associated flows are
  non-generic and typically structurally unstable.}, while for both
$I_{\rm V}=0$ \footnote{It is worth stressing that, what is commonly
  referred to as the vorticity in the context of the two dimensional
  xy model, is actually the Poincar\'e-Hopf index, see e.g.,
  Refs. \cite{Kosterlitz_1973,Kosterlitz_1974}.}.  Similarly to what
occurs for the vorticity, $I_{\rm P}(C)$ does not change
unless a critical point with non zero Poincar\'e-Hopf index (a vortex,
a saddle or an extremum) crosses $C$.

These topological considerations are generic and apply to any system
described by a complex scalar order parameter. The physical
implementation we consider in this Letter consists in the the
propagation of a linearly polarized laser beam of wavelength
$\lambda_0=2\pi/k_0=780$ nm in a cell filled with a nonlinear medium
consisting in a natural Rb vapor at a temperature
$T\approx 120\degree$. Within the paraxial approximation, denoting as
$z$ the coordinate along the beam axis and $\vec{r}$ the transverse
coordinate, this propagation is described by a complex scalar field
$\psi(\vec{r},z)$ which obeys a generalized nonlinear Schr\"odinger
equation \cite{Kivshar-Agrawal} where $z$ plays the role of an
effective time:
\begin{equation}\label{eq.NLS}
 i\, \partial_z \psi
  = - \frac{1}{2 n_0 k_0}
  (\partial_x^2+\partial_y^2) \psi
  + k_0 n_2|\psi|^2\, \psi
  -\frac{ i}{2\, \Lambda_{\rm abs}}\, 
  \psi ,
\end{equation}
$|\psi|^2$ being the intensity, expressed in
W.mm$^{-2}$. $\Lambda_{\rm abs}$ describes the effects of
absorption: if ${\cal T}$ denotes the coefficient of energy transmission, then $\Lambda_{\rm abs}=-z_{\rm max}/\ln({\cal T})$, where $z_{\rm max}=7$ cm is the total length of propagation through the vapor.
$n_0$ is the
refractive index of the medium and $n_2$ is the nonlinear Kerr coefficient.  The values of the parameters are ${\cal T}=0.16$, $n_0=1$ and
$n_2=2.2\times10^{-4}$ W$^{-1}$.mm$^2$ \cite{supplemental}.  

\begin{figure}
\centering
\includegraphics[width=0.9\linewidth]{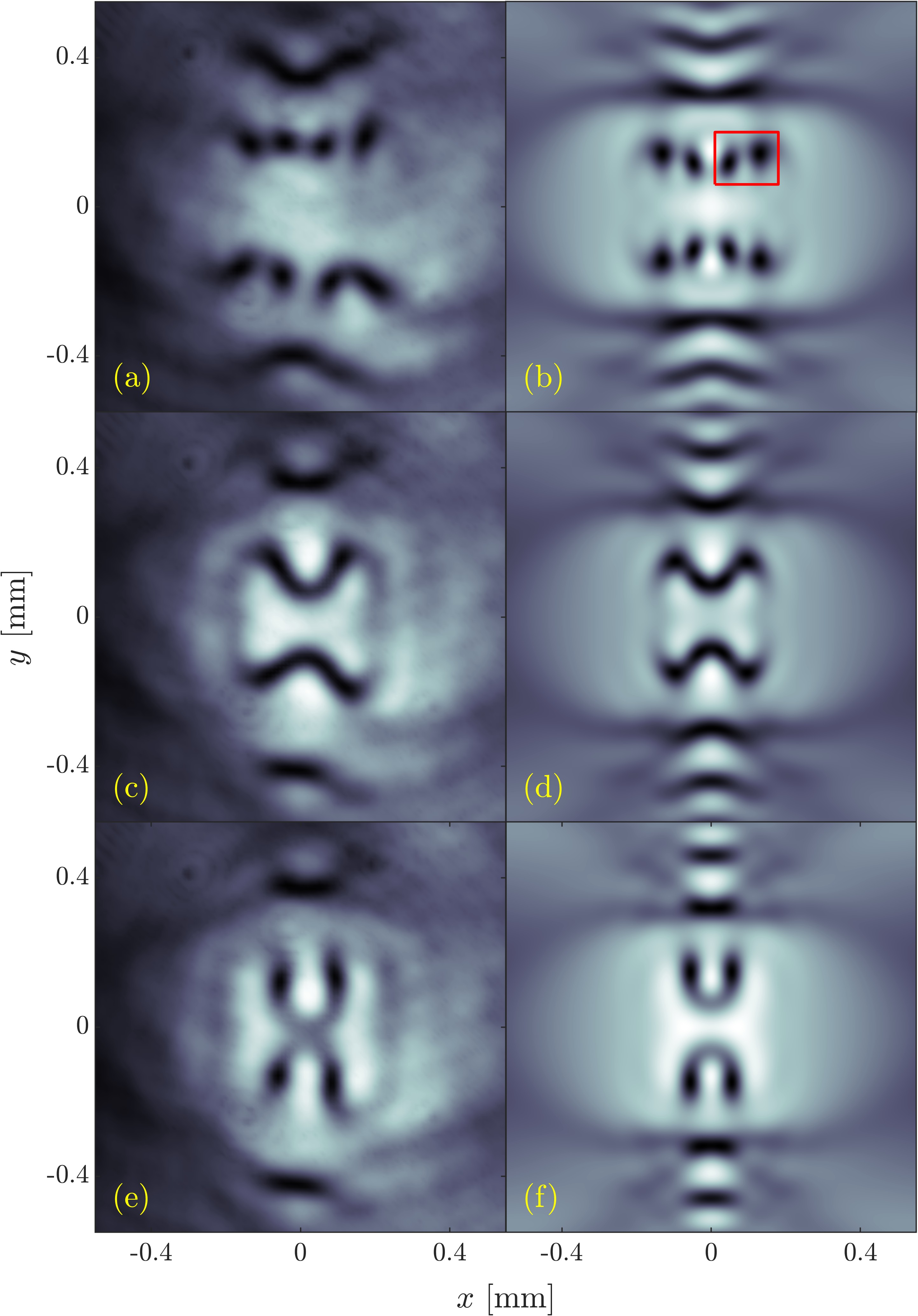}
\caption{Comparison of experimental measurements (left plots) and
  simulations (right plots) of the beam intensity pattern at the exit
  of the vapor. The initial amplitude is given by \eqref{psi_init} with
  $\varPhi_2=0.96\,\pi$ in panels (a) and (b); $\varPhi_2=\pi$ in
  panels (c) and (d) and $\varPhi_2=1.05\,\pi$ in panels (e) and
  (f). The red rectangle in panel (b) marks the location of a
  vortex-antivortex pair whose formation is analysed below, see
  Fig. \ref{fig3}.}\label{fig2}
\end{figure}

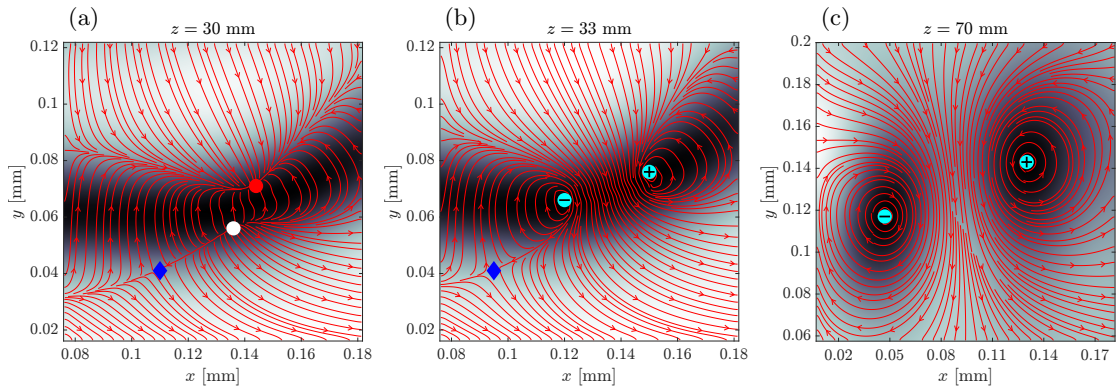
\begin{figure*}
    \centering
    \begin{picture}(15,4.5)   
\put(0,0){\includegraphics[height=0.27\linewidth]{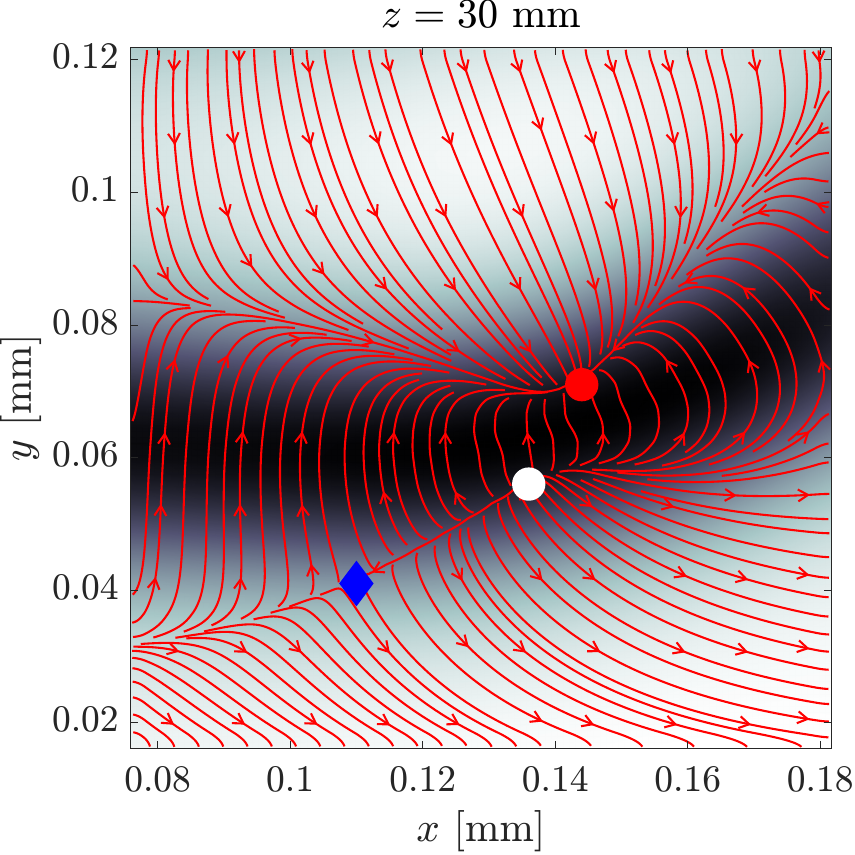}}
\put(5,0){\includegraphics[height=0.27\linewidth]{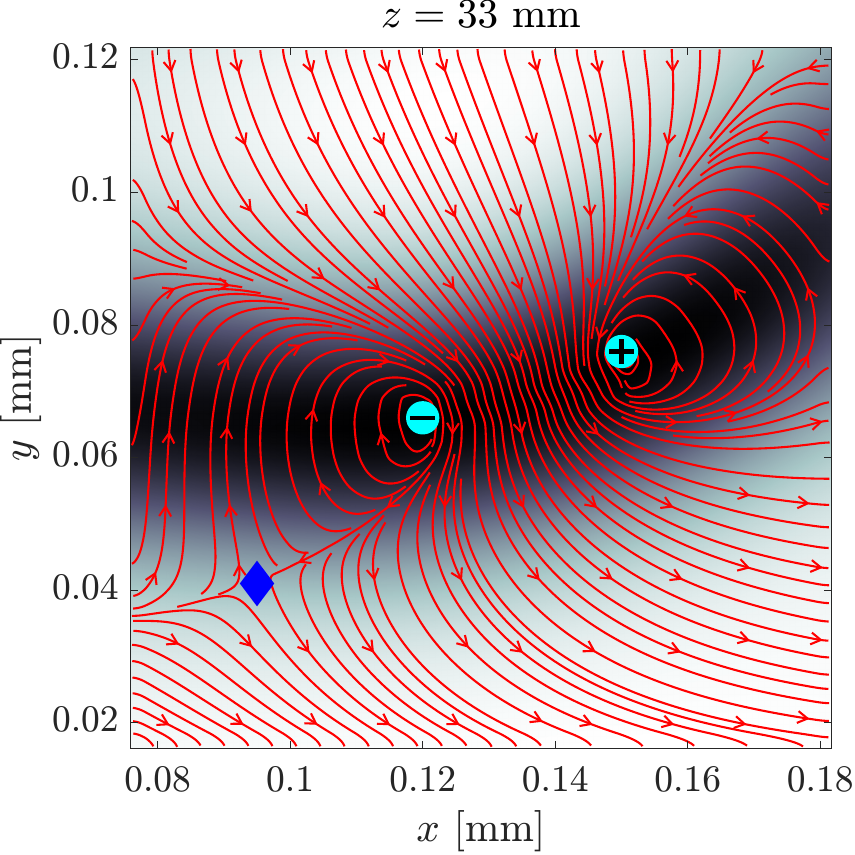}}
\put(10,0){\includegraphics[height=0.27\linewidth]{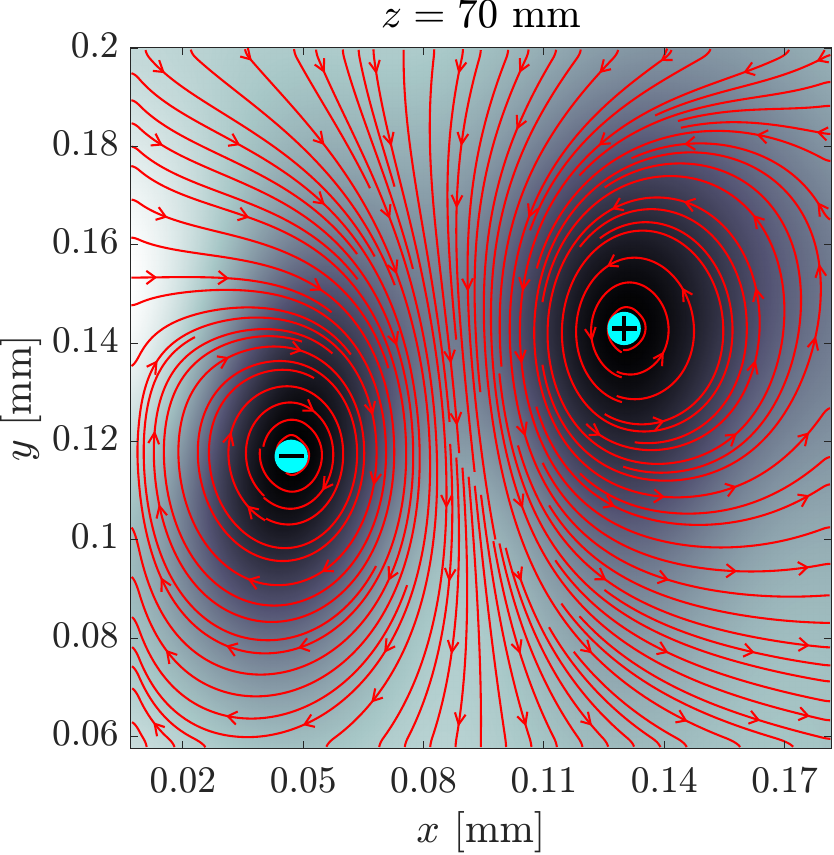}}
\put(0.8,4.8){(a)}
\put(5.8,4.8){(b)}
\put(10.8,4.8){(c)}
\end{picture}
   \caption{Snapshots of simulations of the intensity pattern at several propagation distances within the nonlinear vapor. The initial profile is \eqref{psi_init} with $\varPhi_2=0.96\,\pi$. The corresponding final intensity pattern is represented in Fig. \ref{fig2}(b) of which plot (c) is a blow up. Regions of low intensity are dark. The oriented curves are the streamlines spanned by the vector field $\vec{v}(\vec{r}\,)$. A red (white) circle locates a phase maximum (minimum), i.e., a stable (unstable) node. Cyan circles are vortices. Their vorticity is indicated by a $+$ or $-$ sign ($I_{\rm\sss V}=\pm 1$).
   The blue diamond is a saddle which plays no part in the vortex formation mechanism.}
    \label{fig3}
\end{figure*}

For studying the above discussed topological constrains we use a
specifically designed incident light pattern which consists in the
superposition of a main Gaussian beam (wide and isotropic) with an
auxiliary one, more tightly focused and anisotropic.  The initial
amplitude accordingly reads
\begin{multline}\label{psi_init}
  \psi(\vec{r},0)=  \sqrt{I_1}\exp\left(-\frac{r^2}{w_{\rm\sss G}^2}\right)
 \\+
 \sqrt{I_2}
  \exp\left(-\frac{x^2}{w_x^2}-
  \frac{ y^2}{w_y^2}\right)\exp{i \, \varphi_2(\vec{r}\,)},
\end{multline}
where $r=|\vec{r}\,|$, $w_{\rm\sss G}=1.1$ mm, $w_x=0.55$ mm,
$w_y=0.08$ mm and $I_1= I_2= 0.4$ W.mm$^{-2}$. The initial phase of
the auxiliary beam is
$\varphi_2(\vec{r}\,)=- k_0 r^2/R_2 + \varPhi_2$, where $R_2=-0.5$ m
is the initial curvature of the wavefront of the auxiliary beam and
$\varPhi_2$ is the global phase difference between the auxiliary and
the main beam.  An antiphase relationship ($\varPhi_2=\pi$)
corresponds to an intensity dip induced by the narrow auxiliary beam
on the main one.  We image the beam pattern at the exit of the cell
for different initial phase differences $\varPhi_2=\pi(1 \pm
0.05)$. This is performed thanks to a wave-front sensor which captures
the amplitude and phase of the near field at the output of the
nonlinear medium. As exemplified in Fig.~\ref{fig1} this enables to
simultaneously measure the output optical fluid intensity $|\psi|^2$
and velocity $\vec{v}$.

Fig. \ref{fig2} compares the experimental and theoretical intensity profiles $|\psi(x,y,z_{\rm max})|^2$ at the exit of the cell. In panels (a) and (b) eight vortices distributed symmetrically with respect to the horizontal and vertical axes are observed which have been created during the nonlinear propagation within the cell.
When increasing the initial phase difference $\varPhi_2$ between the main and auxiliary beam, the vortices close to the $y$ axis get even closer [panels (c) and (d)] and eventually merge [panels (e) and (f)].
The agreement between the experimental and numerical results displayed in Fig. \ref{fig2}
is excellent, especially if one considers that there are no free parameters: all the constants of the model have been determined by independent experimental measurements \cite{supplemental}. This
validates the  use of the nonlinear Schr\"odinger equation \eqref{eq.NLS} for studying the intermediate steps ($0<z<z_{\rm max}$) which are not accessible in our experiment.

The dynamics of the critical points during the propagation within the
nonlinear vapor can be complex, but it always fulfills the previously
stated topological requirements. For instance, in numerical
simulations, we have observed the concomitant apparition of a phase
saddle and of a phase extremum, a process which preserves the total
Poincar\'e-Hopf index.  In a similar way, the topological rules impose
that the annihilation of a vortex-antivortex pair be associated to
the simultaneous disappearance of two saddles in order to ensure the
conservation not only of $I_{\rm V}$ but also of $I_{\rm P}$. This is
the process at play in the disappearance of the two pairs of central
vortices observed in Fig. \ref{fig2}, when going from the top to the
bottom row.  We will not go in the particulars of this mechanism here
(see however the discussion in \cite{supplemental}) because it has
been described in detail by the Bristol team \cite{Nye1988} and also
because it is seldom observed in our investigation. In the following
we describe an alternative mechanism of vortex formation, much more
often encountered in our setting: two phase extrema collide and
annihilate one the other, giving birth to a vortex-antivortex
pair. During this process the total Poincar\'e-Hopf index and total
vorticity keep the value 2 and 0, respectively.  This mechanism is at
the origin of the formation of the two vortices in the red square of
Fig. \ref{fig2}(b).  Numerically computed intermediate beam structures
leading to the output pattern shown in Figs. \ref{fig2}(a) and
\ref{fig2}(b) are presented in Fig. \ref{fig3}.  A phase minimum
(white dot) approaches a phase maximum (red dot), pinching a low
density region. The two extrema annihilate each other giving birth to
a vortex-antivortex pair [cyan circles in Fig. \ref{fig3}(b)]. The
fact that the two vortices have opposite vorticity is clearly seen
from the orientation of the streamlines in the vicinity of each of
them. After their formation, the two vortices slowly drift apart,
eventually reaching in Fig. \ref{fig3}(c) the configuration identified
by a red rectangle in Fig. \ref{fig2}(b).

\begin{figure}
\centering
\begin{picture}(8.5,4)
\put(0,0){\includegraphics[width=0.49\linewidth]{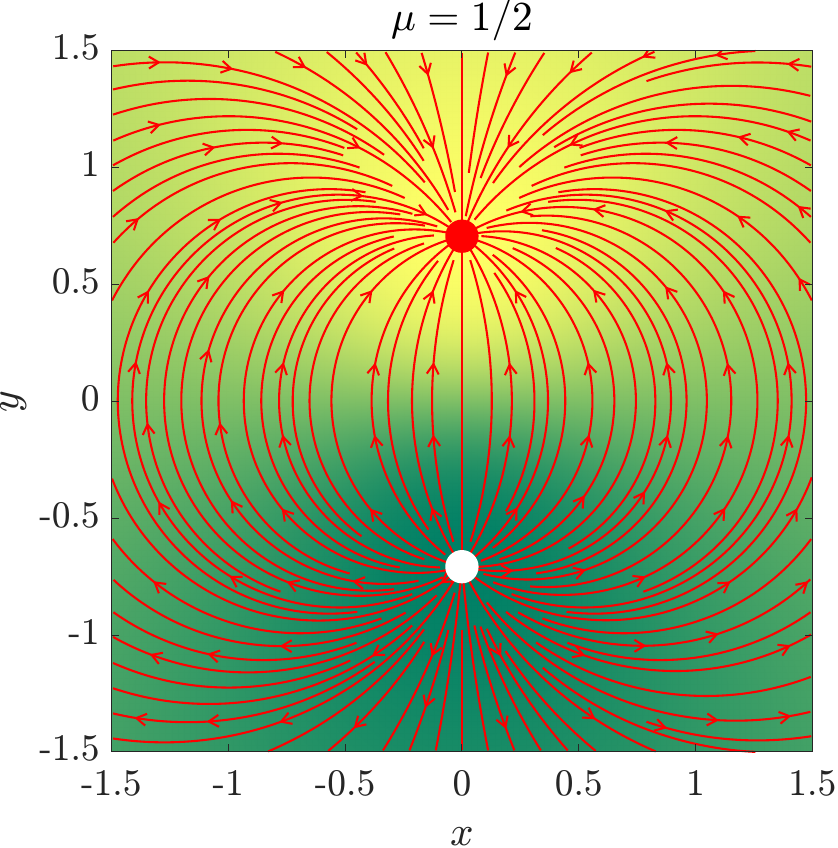}}
\put(4.4,0){\includegraphics[width=0.49\linewidth]{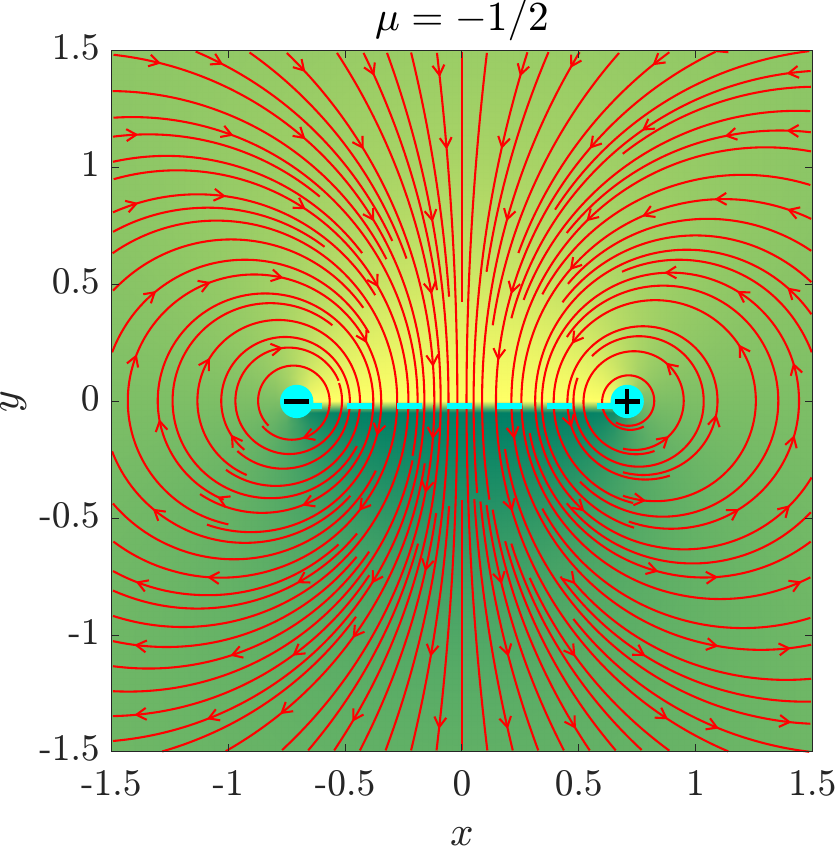}}
\put(0.6,4.2){(a)}
\put(5.0,4.2){(b)}
\end{picture}
\caption{Phase portraits of the dynamical system~\eqref{eq:ds},\eqref{eq:fh}  [or equivalently \eqref{eq:ds},\eqref{eq:S}] with $\sigma=1$ for two different values of the bifurcation parameter $\mu$. The color scale corresponds here to the phase $S_{\rm f\hspace{0.1mm}H}\in [-\pi,\pi]$ (light yellow corresponds to $S_{\rm f\hspace{0.1mm}H}=\pi$ and dark green to $S_{\rm f\hspace{0.1mm}H}=-\pi$). Note that the position of the $2\pi$-jump of $S_{\rm f\hspace{0.1mm}H}(\vec{r}\,)$ [dashed line in panel (b)], is arbitrary and fixed by the choice of constant of integration in~\eqref{eq:S}.} 
\label{fig4}
\end{figure}

The structure of the flow, entailed in the velocity field $\vec{v}(\vec{r}\,)$, can be interpreted within the theory of dynamical systems by considering streamlines (red lines in Figs.~\ref{fig1} 
and~\ref{fig3}) as trajectories of a 2D system:
\begin{equation}
\label{eq:ds}
    \frac{{\rm d}\vec{r}}{{\rm d} \gamma} = \vec{v}(\vec{r}\,),
\end{equation}
with $\gamma$ an arbitrary parametrization on the trajectory. 
In the terminology of dynamical systems, phase extrema are known as nodes (stable or unstable) and saddles as saddle points~\cite{strogatz_nonlinear_2015}. Although vortices are not equilibria of the velocity field, the streamlines encircling a vortex are closed trajectories, and vortices can be seen as ``centers'' of the dynamical system~\eqref{eq:ds}. Within this framework, the change of topology of the flow can be viewed as a bifurcation of~\eqref{eq:ds}: for instance, the above mentioned concomitant apparition of a saddle point (phase saddle) and of a node (phase extremum) is described by a so-called saddle-node bifurcation.
In the same line, the mechanism described previously, and displayed in Fig.~\ref{fig3}, appears in the  fold-Hopf bifurcation \cite{Guckenheimer2002,Kuznetsov2004}
for which a generic normal form is given explicitly in \cite{supplemental}. For the present discussion it suffices to consider the system~\eqref{eq:ds} with the specific form:
\begin{equation}
\label{eq:fh}
    \vec{v} = \vec{v}_{\rm \,f\hspace{0.1mm}H}(\vec{r}\,) \equiv
        -2\sigma x y \, \vec{e}_x + (\mu + \sigma x^2- y^2) \, \vec{e}_y,
\end{equation}
where $\sigma=\pm 1$ is fixed, and $\mu\in\mathbb{R}$ is a parameter of the bifurcation. The phase portrait of the dynamical system~\eqref{eq:ds},\eqref{eq:fh} for $\sigma=1$ and
two different values of $\mu$ (before and after the bifurcation) is shown in Fig.~\ref{fig4}. 
In this case, the stable and unstable nodes (red and white dot respectively) which exist when $\mu>0$ annihilate when $\mu$ becomes negative to form two centers (represented by cyan circles); note that the latter are not singularities but true equilibria of the velocity field~\eqref{eq:fh}. 

The velocity field \eqref{eq:fh}
is not that of a potential flow, as should be the case for a quantum fluid. However, it is possible  to derive a 
potential flow
which shares the same phase portrait.
The corresponding velocity field reads (see \cite{supplemental})
\begin{equation}
\label{eq:S}
\vec{v} = \vec{\nabla} S_{\rm f\hspace{0.1mm}H},\quad S_{\rm f\hspace{0.1mm}H}(\vec{r}\,)\equiv\arg\big[x^2+\sigma(y^2+\mu)+ i \sigma y\big].
\end{equation}
The system \eqref{eq:ds},\eqref{eq:S} is not only a gradient flow, it
also obeys the Onsager-Feynman quantization condition
\cite{note_OF}. In particular the centers
of~\eqref{eq:ds},\eqref{eq:fh} are replaced by singularities (where
$S_{\rm f\hspace{0.1mm}H}$ is ill-defined) that are encircled by
closed orbits along which the circulation of
$\vec{\nabla} S_{\rm f\hspace{0.1mm}H}$ is $\pm 2\pi$ [as depicted in
Fig.~\ref{fig4}(b)], i.e., quantum vortices. Note that
$S_{\rm f\hspace{0.1mm}H}$ is not the phase of a wave-function which
exactly obeys the nonlinear Schr\"odinger equation
\eqref{eq.NLS}. However, comparing the phase portraits of
Fig.~\ref{fig4} with the flow patterns obtained in Figs.~\ref{fig3}(a)
and \ref{fig3}(b) shows that varying $\mu$ in \eqref{eq:S} effectively
reproduces the local flow pattern of a $z$-varying wave-function
solving \eqref{eq.NLS}.  Besides, $S_{\rm f\hspace{0.1mm}H}$ fulfills
the requirements expected from the phase of the order parameter
\eqref{eq:psi} of a 2D quantum fluid.

It is interesting to remark that the normal form \eqref{eq:fh}, once
modified to derive from the velocity potential \eqref{eq:S} as just
explained, also describes when $\sigma=-1$ the scenario of vortex
annihilation presented in~\cite{Nye1988} which we henceforth denote as
the Bristol mechanism: two vortices and two saddle points annihilate
when $\mu$ goes from positive to negative, yielding a featureless
flow. Notably, the model wavefunction given in~\cite{Nye1988} reduces
to $\psi = x^2-y^2- \mu- i y$ close to the bifurcation point (see
\cite{supplemental}), i.e., its phase is $S_{\rm f\hspace{0.1mm}H}$
with $\sigma=-1$, validating the analogy presented here: the normal
form of the fold-Hopf bifurcation provides an approximated theoretical
model of the Bristol mechanism.

The mechanism of vortex formation illustrated in Figs.~\ref{fig3} and
\ref{fig4}, although generic, cannot be observed in the special case
of an incompressible 2D quantum fluid, such as commonly
used to model liquid helium films for instance. Indeed, in such a
system the phase $S$ is a harmonic function, which, by the maximum
principle cannot have maxima nor minima: the only possible critical
points with zero velocity are saddles and no phase extrema occur,
contrary to what is observed in Fig. \ref{fig3} (see an extended
discussion of this point in \cite{supplemental}).  Phase extrema are
also forbidden in a stationary (i.e., $z$-independent in our case)
system, as proven in Ref. \cite{Nye1988}, but nothing prevents their
formation in a $z$-dependent configuration. Indeed, such extrema have
been theoretically considered \cite{Freund1995} and experimentally
observed in a random linear speckle pattern \cite{Shvartsman1995}, but
were found to be relatively scarce, being outnumbered in a ratio 14:1
by saddles.  Although our use of a specific initial condition
\eqref{psi_init} prevents a systematic statistical study, we also observe that
phase extrema are less numerous than saddles. This corresponds to
physical intuition: extrema are typically born in saddle-node
bifurcations which create an equal number of extrema and saddles,
whereas pairs of saddles could be additionally created thanks to the
Bristol mechanism.  More significantly, the new mechanism of vortex
formation we have identified and observed in many instances,
efficiently diminishes the number of extrema. As a result, when
vortices proliferate, saddles tend to be more numerous than extrema.

In conclusion we emphasize that our experiment uses a new generation
of optical techniques which enable a precise measure of both the
intensity and the phase of a light sheet
\cite{Nardin2010,Anton2012,Dominici2015,Sitnik2022,Panico2023,Abobaker2022}. As
demonstrated in the present Letter, this offers the possibility of an
accurate and simple location not only of vortices but also of other
critical points, such as saddles. This enabled us to obtain evidences
of several (topologically constrained) mechanisms of formation of vortices
and of associated singular points  in the time domain, with an account
of the evolution of the streamlines. As far as vortex formation is
concerned, we experimentally demonstrated a scenario proposed more
than 30 years ago (the Bristol mechanism). We also identified a new
scenario, simpler and more common in our setting, in which two nodes
collide and give birth to a vortex-antivortex pair.  This process
requires a non stationary flow and a compressible fluid.  We showed
that the two mechanisms of vortex formation (Bristol and nodes
collision) pertain to the same fold-Hopf type of bifurcation. We
demonstrated that the corresponding normal form can be enriched in
order to account for the quantum nature of our system. This suggests
that these mechanisms are universal. It would thus be of great
interest to uncover to what extent they are involved in the nucleation
or annihilation of vortices and of more exotic defects recently
studied in Refs. \cite{Seo2016,Kang2019,Kwon2021,Richaud2023} or also
during the Kibble-Zurek process \cite{Chomaz2015,Ko2019}.  As a final
remark we stress that our study illustrates the efficiency of tools
issued from the theory of dynamical systems to investigate the route
to turbulence. This opens the path of a new line of research devoted
to the statistical study of nodes and saddles dynamics in a turbulent
quantum fluid.

\begin{acknowledgments}
TC and NP would like to thank the Isaac Newton Institute (INI) for Mathematical Sciences for support and hospitality during the programme ``Dispersive hydrodynamics: mathematics, simulation and experiments, with applications in nonlinear waves''  when part of the work on this Letter was undertaken. The work of TC was partially supported by the Simons Fellowships during the INI Programme. 
\end{acknowledgments}

\bibliography{biblio}

\begin{thebibliography}{94}
\bibitem [{not()}]{note_OF}%
  \BibitemOpen
  \href@noop {} {}\bibinfo {note} {The fact that $I_{\rm V}$ is an integer is
  known as the Onsager-Feynman quantization condition, see
  \cite{Onsager1949,Feynman1955}}\BibitemShut {NoStop}%
\end{thebibliography}

\end{document}


\title{Supplemental material to : Topological Constraints on the Dynamics of Vortex Formation in a Two-dimensional Quantum Fluid}

\author{T. Congy}
\affiliation{Department of Mathematics, Physics and Electrical Engineering, Northumbria University, Newcastle upon Tyne NE1 8ST, United Kingdom}
\author{P. Azam}
\author{R. Kaiser}
\affiliation{Institut de Physique de Nice, Université Côte d'Azur, CNRS, F-06560 Valbonne, France}
\author{N. Pavloff}
\affiliation{Universit\'e Paris-Saclay, CNRS, LPTMS, 91405, Orsay, France}
\affiliation{Institut Universitaire de France (IUF)}
\maketitle

\section{Experimental setup}

\begin{figure}
    \centering
    \includegraphics[width=\linewidth]{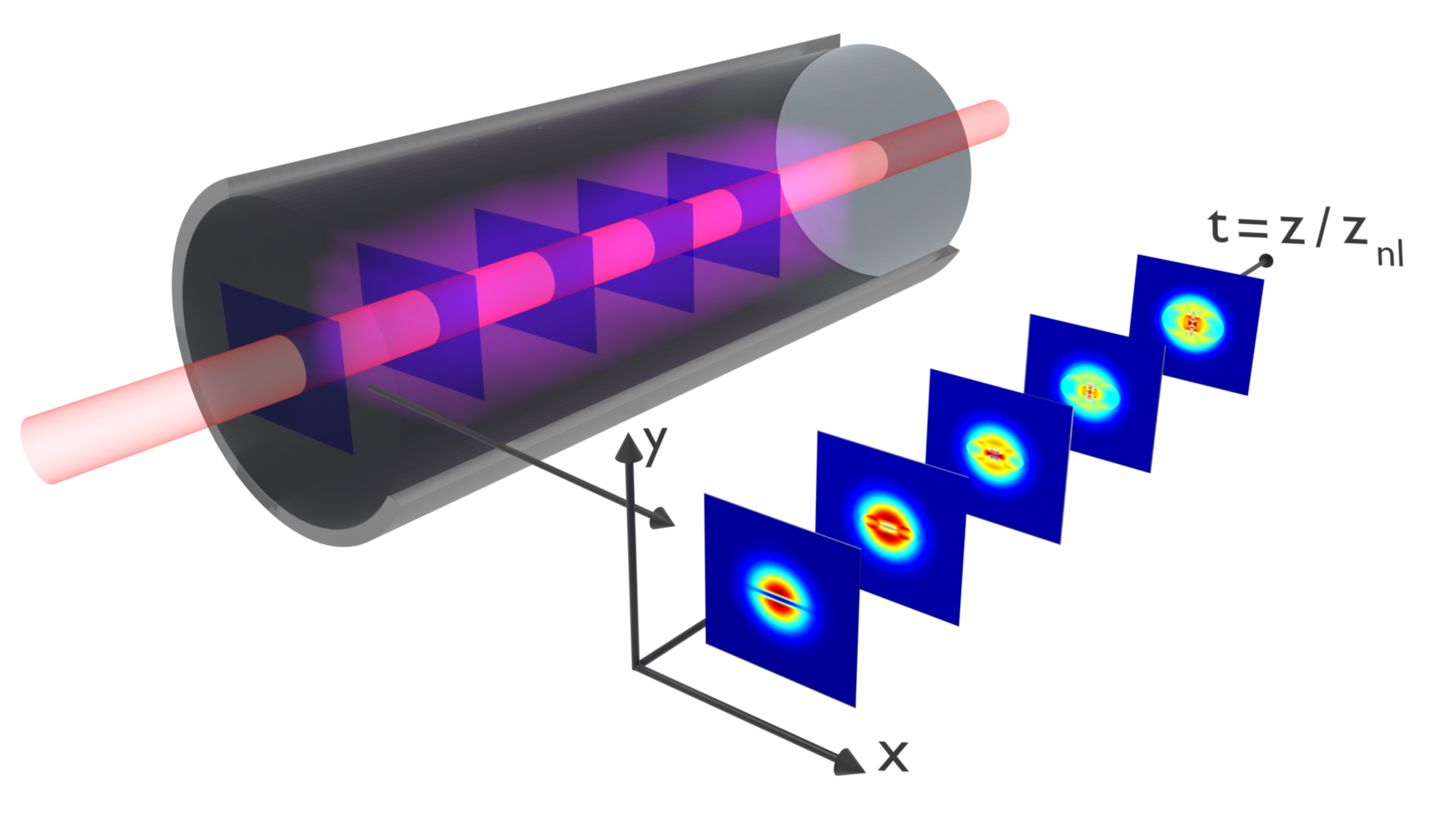}
    \caption{Sketch of the propagation of the beam within the nonlinear medium. The side pictures depict the transverse intensity pattern at different propagation distances. The quantity $z_{\rm nl}$ is the nonlinear length defined in Eq. \eqref{znl_xi}.} 
    \label{fig:S1}
\end{figure}

In this section we describe the experimental setup and discuss some of its characteristics.\\
As discussed in the main text and sketched in Fig. \ref{fig:S1}, the experimental system consists in the propagation of a laser beam ($\lambda_0=780$ nm) aligned with the axis $Oz$ through a cell of length $z_{\rm max}=7$ cm filled with a hot atomic vapor (natural isotopic mixture of Rubidium). Such a vapor behaves as a Kerr nonlinear medium whose nonlinearity can be tuned {\it via} the frequency detuning of the beam with respect of the D$_2$ line of Rb, the beam intensity and the atomic density of the vapor within the cell (monitored through the temperature of the vapor).

\begin{figure}
    \centering
    \includegraphics[width=\linewidth]{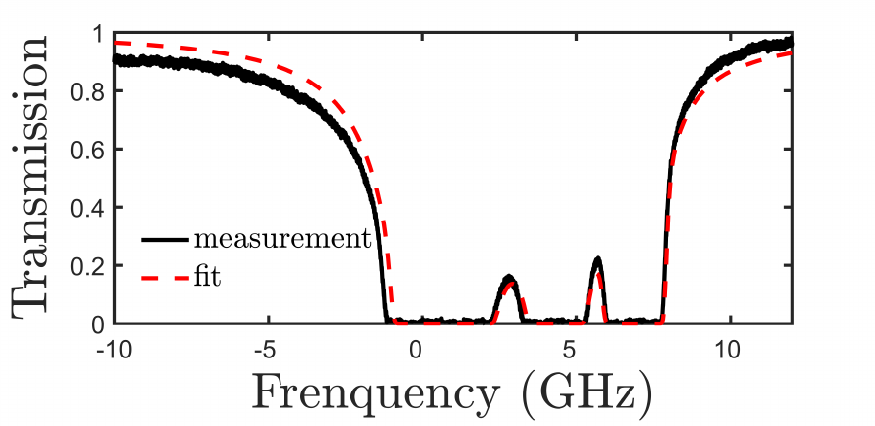}
    \caption{Transmission spectrum for a vapor temperature $T= 120\degree$. The frequency range, of about 25 GHz, is determined by the scanning laser. The frequency is measured with respect to the D$_2$ spectral line of $^{87}$Rb.
    The beam is red shifted with respect to this reference, with a frequency detuning $\Delta = - 0.75$ GHz. The corresponding measured transmission coefficient is ${\cal T}=0.16$.}
    \label{fig:S2}
\end{figure}

Note that to extract atomic gas temperature, we measure the transmission profile of a weak laser beam after propagation through the vapor as a function of the laser detuning. Fitting this data (as shown in Fig. \ref{fig:S2}) with a numerical simulation taking into account atomic lines of both isotopes, rubidium vapor pressure as a function of the temperature \cite{siddons2008absolute,agha2011time,Baudouin14} and the Doppler broadening, we can deduce the atomic density and therefore link it to the gas temperature using the vapor pressure \cite{steck} and ideal gas law.

\begin{figure*}
    \centering
    \includegraphics[width=0.8\textwidth]{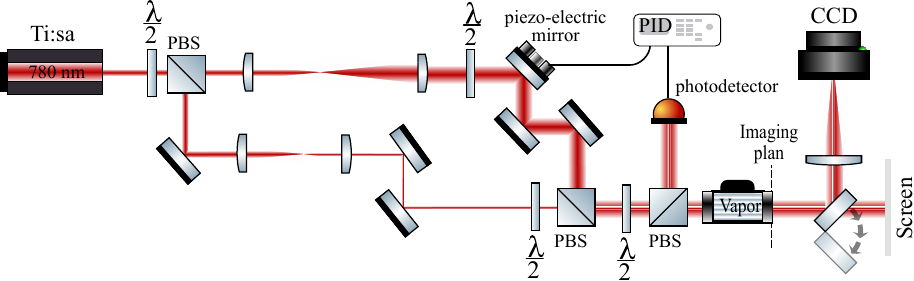}
    \caption{Sketch of the experimental apparatus. A $780$ nm laser beam is splitted via a Mack-Zender interferometer, each arm being separately shaped and then overlapped to create the initial condition: a wide gaussian beam (background) in destructive interference with an elliptical gaussian beam. A PID linked to a piezo-electric mirror controls and locks the relative phase between both beams. After passing through the atomic vapor, the beam at the output of the cell is imaged on a CCD camera. The far-field can also be observed on a screen to measure the nonlinear phase shift.}
    \label{fig:S3}
\end{figure*}

We reproduce here for legibility the equation governing the propagation of the beam's complex amplitude $\psi(\vec{r},z)$  within the nonlinear vapor:
\begin{equation}\label{eq.NLS}
 i\, \partial_z \psi
  = - \frac{1}{2 n_0 k_0} (\partial_x^2+\partial_y^2) \psi
  + k_0 n_2|\psi|^2\, \psi
  -\frac{ i}{2\, \Lambda_{\rm abs}}\, 
  \psi ,
\end{equation}
where $z$ is the longitudinal coordinate and the transverse coordinate is $\vec{r}=x\,\vec{e}_x+y\,\vec{e}_y$, see Fig. \ref{fig:S1}. In this equation
$n_0=1$ is the refractive index of the medium
and $\Lambda_{\rm abs}=-z_{\rm max}/\ln({\cal T})$ phenomenologically describes the effects of
absorption (${\cal T}=0.16$ is the transmission coefficient, see Fig. \ref{fig:S2}). The coefficient $n_2$
is the nonlinear index of refraction. It can be modified by changing the detuning between the frequency of the laser and that of the D$_2$ line of Rb, or the atomic density of the vapor (i.e., the temperature). In our study we use a red detuned laser; in this condition $n_2>0$ and the nonlinear medium has self-defocusing properties.
The nonlinear index of refraction can be determined by an auxiliary experiment \cite{Wu2020} which consists in measuring the nonlinear phase shift 
\begin{equation}\label{eq.NL.phase}
    \Delta\phi_{\rm nl}=k_0 z_{\rm max} n_2 I
\end{equation}
acquired by a gaussian beam of intensity $I$ during its propagation through the nonlinear medium.
This technique, based on the self-phase modulation effect, consists in imaging the far field intensity after propagation in the vapor. Every $2\pi$ phase shift accumulated will appear as a ring in the intensity profile; counting the number of concentric rings gives a good estimate of the nonlinear phase shift and therefore of the Kerr coefficient $n_2$.
The study presented in the main text has been performed for $ \Delta\phi_{\rm nl}=50$ ($\pm\pi$) which corresponds to 
$n_2=2.2(2)\times10^{-4}$ W$^{-1}$.mm$^2$. Experiments were generally performed with an atomic vapor of density $\rho_{at}\approx 2\times10^{19}$ atoms/m$^{3}$ (at a temperature of $T\approx 120$\degree), and a laser detuning varying from $-10$ GHz to $-1$ GHz with respect to the $^{87}$Rb D$_2$ transition $5S_{1/2}(F=2)- 5P_{3/2}(F=1,2,3)$.

The optical field used for the study presented in the main text is composed of a gaussian background beam (with waist $w_{\rm\sss G}=1.1$ mm and power $P_{\rm\sss G}=800$ mW) overlapped with an elliptical gaussian beam (with dimensions $w_{x}=550$ $\mu$m and $w_{y}=80$ $\mu$m) having a central intensity equal to that of the background beam, cf. Eq.~\eqref{psi_init_adim}. 
The interference of these two beams is set as initial condition. Modifying the global relative phase between the two beams, or the radius of curvature of the elliptical beam, or titling the elliptic beam with respect to the background one, leads to modifications of the initial velocity pattern of the fluid of light allowing to frame the dynamics of the system while keeping the parameters of the nonlinear Schr\"odinger equation \eqref{eq.NLS} fixed.
The main power of the field is sent to the rubidium vapor while a small part is sent to a photo-detector linked to a proportional–integral–derivative controller (PID) as depicted in Fig. \ref{fig:S3}. This allows to lock and scan the relative phase $\varphi_2(\vec{r}\,)$ between the beams via a piezo-electric mirror installed on the interferometer (on the arm of the background beam), offering a precise and stable control of the initial fluid's velocity \cite{Azam2022}.

\begin{figure}
\centering
\begin{picture}(8.5,8.9)
\put(0,0){\includegraphics[width=0.99\linewidth]{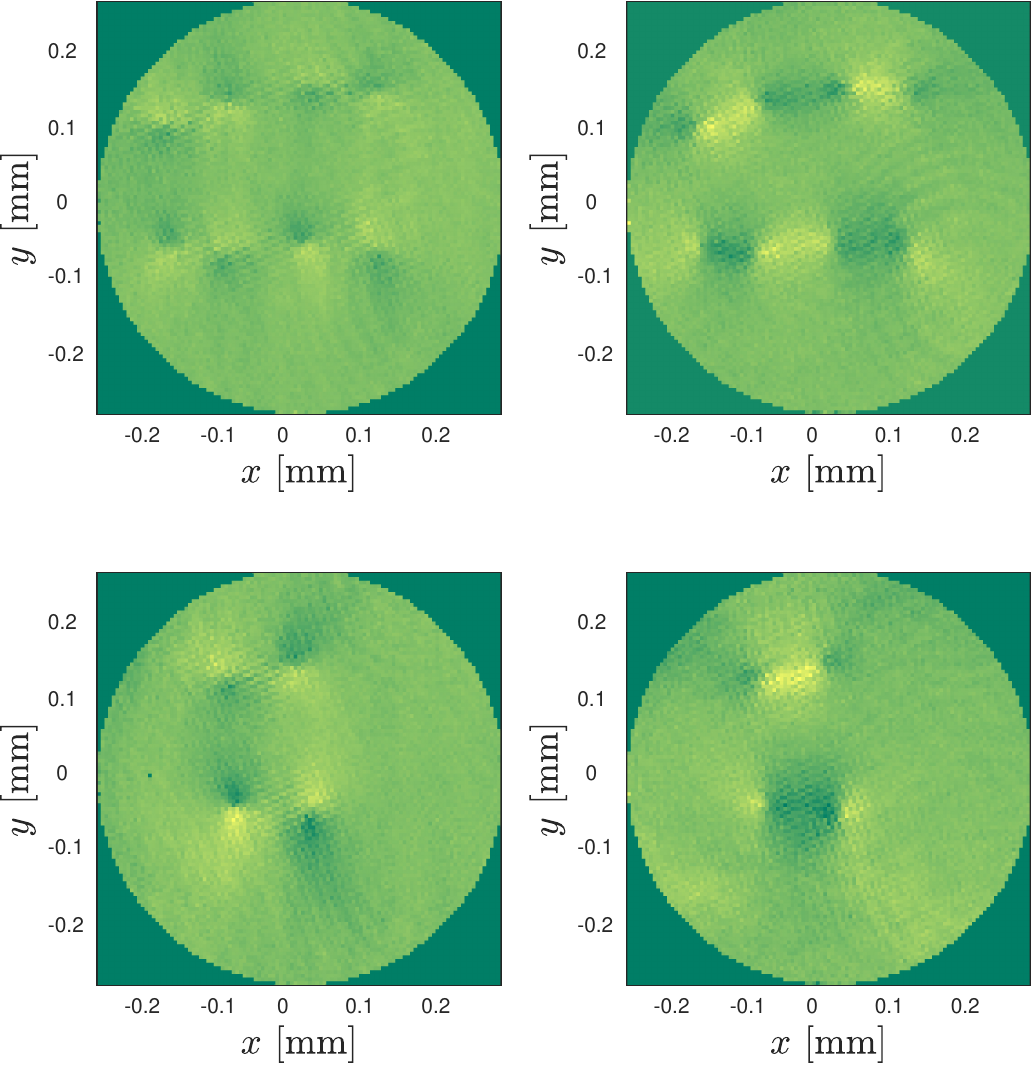}}
\put(0.85,5.55){\bf\color{White}(a)}
\put(5.25,5.55){\bf\color{White}(b)}
\put(0.85,0.8){\bf\color{White}(c)}
\put(5.25,0.8){\bf\color{White}(d)}
\end{picture}
\caption{Left (right) column: Coordinates of the phase gradients along the $x$ ($y$) axis as measured at the output of the nonlinear cell. These are used to construct the streamline patterns presented in Fig. \ref{fig:S4}. Panels (a) and (b)
are the phase gradients corresponding to Fig. \ref{fig:S4}(a). Panels (c) and (d) correspond to Fig. \ref{fig:S4}(c).} 
\label{fig:PhaseGradient}
\end{figure}

In this context, the transverse velocity of the fluid of light can be written as $\vec{v}\propto (\partial_x S \, \vec{e}_x + \partial_y S\, \vec{e}_y)$ where $S(\vec{r},z)$ is the phase of the electromagnetic field. The use of a Phasics camera has been necessary for the experimental observation of the fluid's flow. Such a device uses a technique based on the quadri-wave lateral shearing interferometry (QWLSI) to locally measure both the intensity and the phase gradient of a beam of light. This tools usually used for adaptative optics offers us a direct access to both the density and velocity of our fluid of light. The wavefront sensor measures the gradient of the phase along the two transverse axis, as reported in Fig.~\ref{fig:PhaseGradient}. From these records we can directly construct the velocity field and then the corresponding streamlines, as displayed in Fig. 1 of the paper and in Fig. \ref{fig:S4} below.

\begin{figure*}
    \centering
    \begin{picture}(17,5)
\put(0,0){\includegraphics[width=0.32\linewidth]{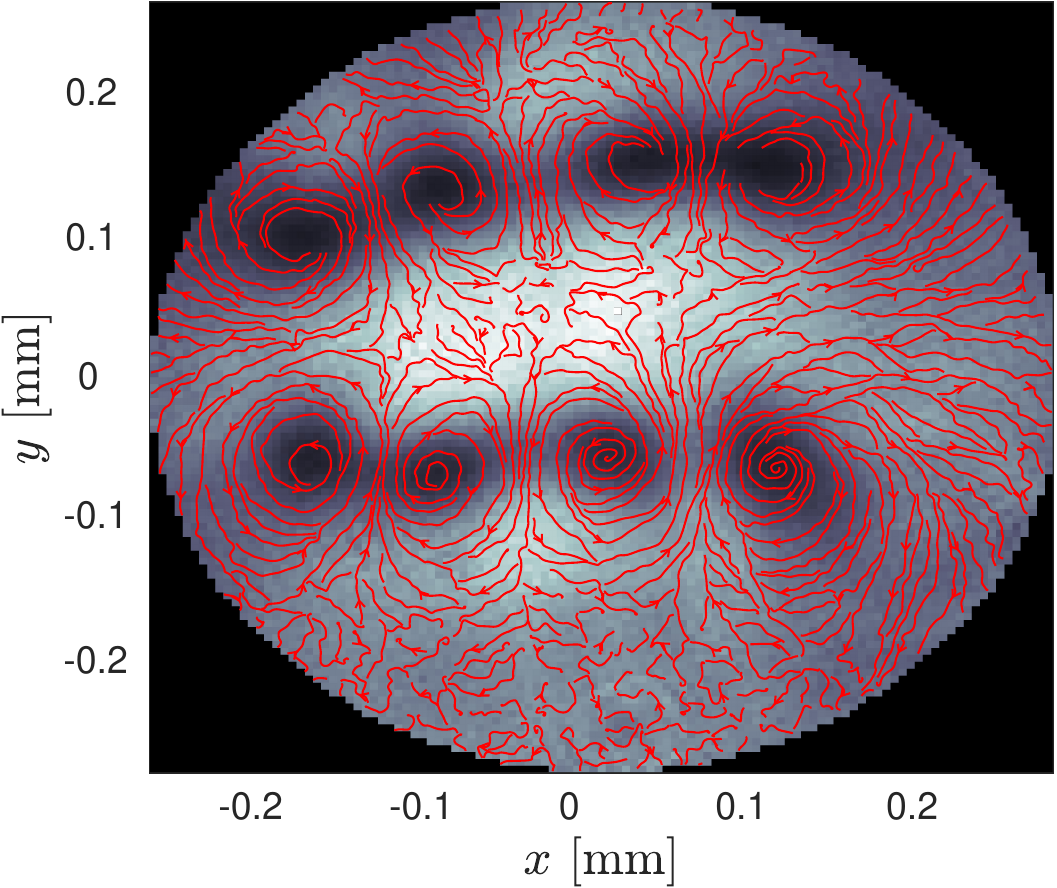}}
\put(5.8,0){\includegraphics[width=0.32\linewidth]{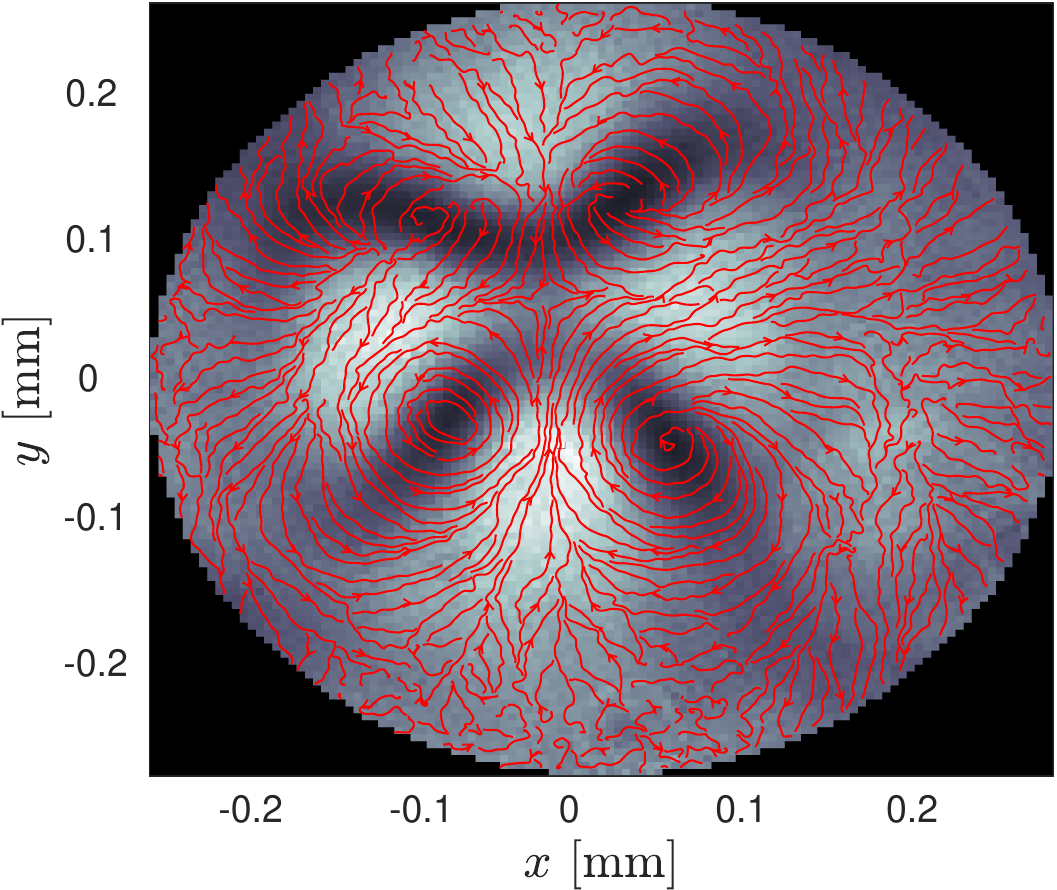}}
\put(11.6,0){\includegraphics[width=0.32\linewidth]{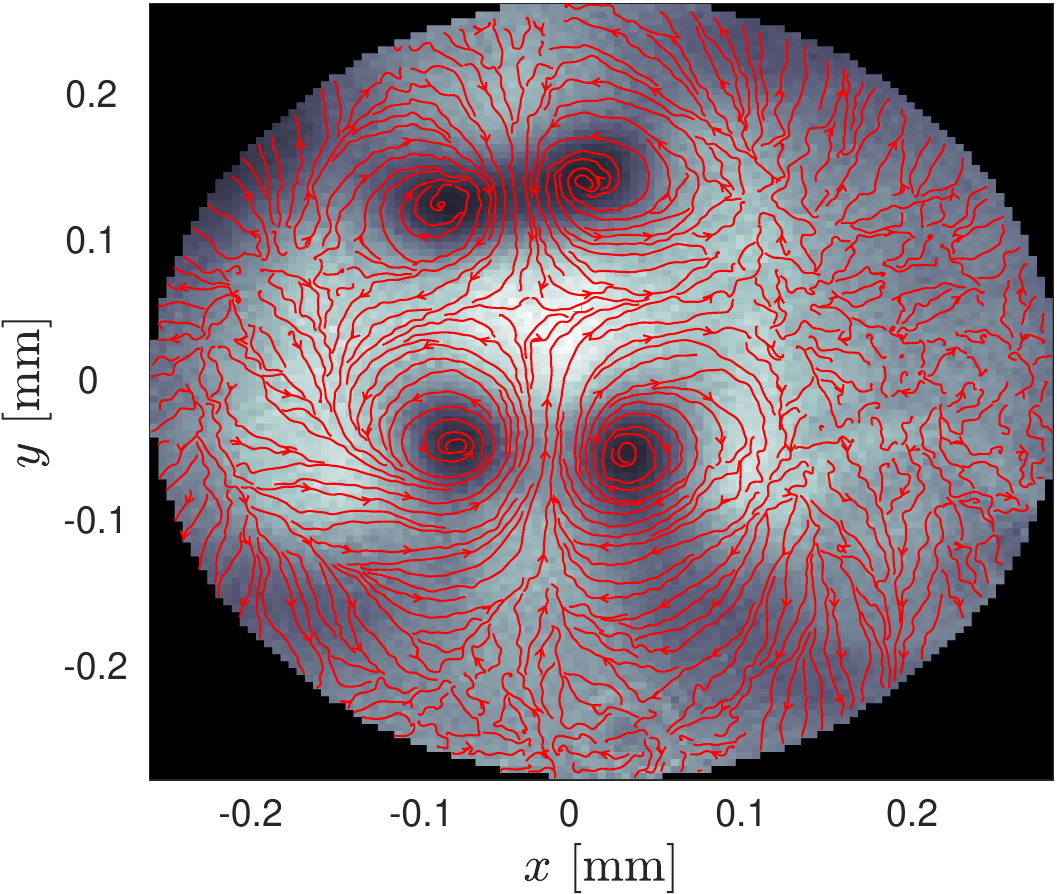}}
\put(0.9,0.9){\color{yellow}\large (a)}
\put(6.7,0.9){\color{yellow}\large (b)}
 \put(12.5,0.9){\color{yellow}\large (c)}
   \end{picture}
    \caption{Measured output intensity profile and streamlines (in red) for different values of the input global phase difference between the main and the elliptical beam: panel (a) $\varPhi_2=0.96\,\pi$; panel (b) $\varPhi_2=\pi$; panel (c) $\varPhi_2=1.05\,\pi$. Regions of lower (higher) intensity are dark (bright). The nonlinear parameter stays fixed ($\Delta\phi_{\rm nl}=50$) during the study, allowing to observe the fluid's evolution in the exactly same conditions.}
    \label{fig:S4}
\end{figure*}

\section{Parameters and results}

Eq. \eqref{eq.NLS} can be put in a non dimensional form by choosing a reference intensity  $I_{\rm ref}$. A natural choice is to take $I_{\rm ref}=I_1$, where $I_1$ is the common intensity of the main and of the auxiliary beam, but other choices are possible. Defining the nonlinear length $z_{\rm nl}$ and the transverse healing length $\xi$ by 
\begin{equation}
\label{znl_xi}
z_{\rm nl}=(k_0 n_2 I_{\rm ref})^{-1}=n_0 k_0 \xi^2,
\end{equation}
and considering the dimensionless quantities $t=z/z_{\rm nl}$, $X=x/\xi$, $Y=y/\xi$ and $\Psi=\psi/\sqrt{I_{\rm ref}}$ makes it possible to write  \eqref{eq.NLS} under the form
\begin{equation}\label{adim.NLS}
    i\partial_t \Psi = \tfrac{1}{2} (\partial^2_X+\partial_Y^2)\Psi + |\Psi|^2 \Psi -i\gamma \Psi\; ,
\end{equation}
where $\gamma=-\tfrac12 \ln{\cal T} \times z_{\rm nl}/z_{\rm max}$. Taking $I_{\rm ref}=I_1=0.4$ W.mm$^{-2}$ and $n_2=2.22\times10^{-4}$ W$^{-1}$.mm$^2$ yields $\xi=13.2$ $\mu$m, $z_{\rm nl}=1.40$ mm and $t_{\rm max}=z_{\rm max}/z_{\rm nl}=50$. 
The value ${\cal T}=0.16$ of the transmission corresponds to  $\gamma=0.018$.
In terms of the dimensionless parameters the nonlinear phase shift 
\eqref{eq.NL.phase} reads $\Delta\phi_{\rm nl}=t_{\rm max} \cdot (I_1/I_{\rm ref})$.

We solve numerically~\eqref{adim.NLS} with the initial profile
\begin{multline}\label{psi_init_adim}
  \Psi(X,Y,0)=  \exp\left(-\frac{R^2}{W_{\rm\sss G}^2}\right)
 \\+
  \exp\left(-\frac{X^2}{W_x^2}-
  \frac{ Y^2}{W_y^2}\right)\exp{i \, \varphi_2(X,Y)},
\end{multline}
where $R^2=X^2+Y^2$. The coefficient $W_{\rm\sss G}=83.3$, $W_x=41.7$ and $W_y=6.06$ are the dimensionless waists of the two beams (for instance $W_{\rm\sss G}=w_{\rm\sss G}/\xi$, where the value of $w_{\rm\sss G}$ is given in the previous section). The phase of the secondary beam reads, in dimensionless coordinates
\begin{equation}
    \varphi_2(X,Y)=-\alpha R^2 + \varPhi_2
\end{equation}
where $\alpha=-2.81\times 10^{-3}$ characterizes the curvature of the wavefront of the auxiliary beam and $\varPhi_2$ is the global phase difference between the main and the auxiliary beam. 

The intensity profiles at $t=t_{\rm max}$ (i.e., at the exit of the nonlinear cell)  are displayed in Figure~2 of the main text for different values of the global phase difference $\varPhi_2$.\\
Fig. \ref{fig:S4} represents an equivalent set of data using the wavefront sensor which gives an experimental access to the streamlines. These data are reconstructed from the phase gradients measured by the wavefront sensor and shown in Fig. \ref{fig:PhaseGradient}.

\subsection{Bristol mechanism}

The field pattern at the exit of the nonlinear cell is represented in Fig. \ref{fig:S4} for different values of the control parameter $\varPhi_2$. We plot not only the intensity but also the streamlines of the fluid of light, which is made possible by our recording of the phase of the field.
By a detailed inspection of this figure and also by using additional numerical simulations for intermediates values of $\varPhi_2$, one can verify that, when increasing the parameter $\varPhi_2$ from the value $0.96\,\pi$, some of the 8 vortices present at the output of the cell in Fig. \ref{fig:S4}(a) merge according to what we call in the main text the ``Bristol mechanism'', for eventually leading to a 4-vortices configuration when $\varPhi_2=1.05\,\pi$, as shown in Fig. \ref{fig:S4}(c).

In this section, instead of studying the influence of a modification 
of the initial phase difference between the two beams on the output intensity pattern, we rather fix the value $\varPhi_2=1.05\,\pi$ and numerically study the behavior of the laser beam within the nonlinear cell (i.e., as a function of $z$). We will see that the general trend is similar, indicating that, as suggested by intuition, modifying the global phase difference between the two beams induces a change of relative velocity which effectively speeds up (or slows down) the nonlinear flow, enabling, by varying $\varPhi_2$,  to observe at the output of the cell, patterns occurring for fixed $\varPhi_2$ at the interior of the cell (which are not observable experimentally).

\begin{figure}
    \centering
    \includegraphics[width=0.49\linewidth]{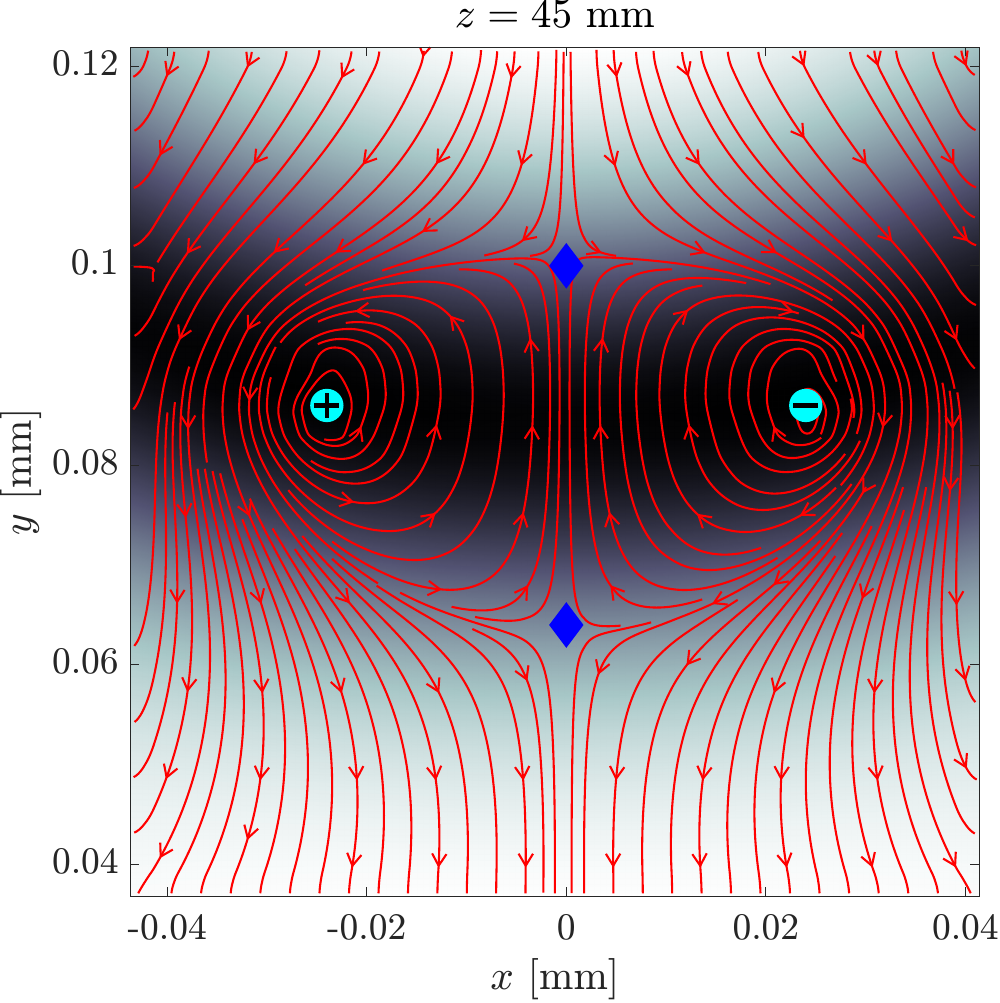}
    \includegraphics[width=0.49\linewidth]{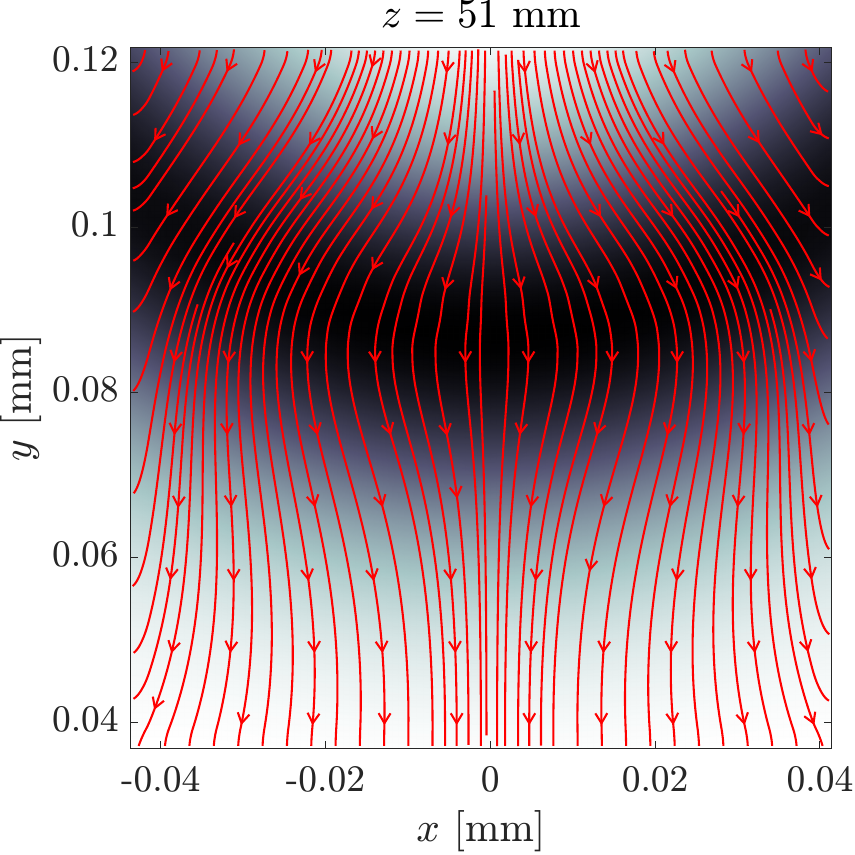}
    \caption{Snapshots of simulations of the intensity pattern at two propagation distances through the nonlinear vapor ($z=45$ mm and 51 mm). The initial profile is \eqref{psi_init_adim} with $\varPhi_2=1.05\,\pi$. Regions of low intensity are dark. The oriented curves are the streamlines spanned by the velocity field $\vec{v}(\vec{r}\,)$. Cyan circles are vortices.  Their vorticity is indicated by a $+$ or $-$ sign ($I_{\rm\sss V}=\pm 1$). Blue diamonds are saddles.} 
    \label{fig:S5}
\end{figure}

If $\varPhi_2 = 1.05\,\pi$, the profile at the exit of the nonlinear cell displays 4 distinct vortices, as shown in Figs. 2(e) and 2(f) of the main text and in Fig. \ref{fig:S4}(c). For this value of $\varPhi_2$, numerical simulations show that 8 vortices are initially generated within the cell.  In short, the mechanism of vortex formation is the following: since $\varPhi_2$ is close to $\pi$, the initial profile displays a region a low density elongated along the $x$ axis.
At the entrance of the nonlinear medium the effective photon-photon interaction suddenly sets in, as described in Ref. \cite{Glorieux2018} which considers a setting similar to ours.
The initial depleted region is accordingly separated in two solitons  counter propagating in the $y$-direction. These two solitons decay  {\it via} the transverse instability (also known as snake instability, see e.g., Refs. \cite{Kuznetsov1988,Tikhonenko:96}) creating a rather large number of vortices (4 for each soliton). We numerically observe that these vortices are created in pairs, through the collision of two nodes (fold-Hopf bifurcation with $\sigma=1$, see below).
During the subsequent propagation within the cell,
among the 8 vortices  two vortex-antivortex pairs annihilate via the Bristol mechanism~\cite{Nye1988}. This annihilation is depicted in Fig.~\ref{fig:S5}: two vortices and two saddles get closer (left plot) and then annihilate during the propagation along the $z$-axis (right plot) \footnote{In the simulations presented in this paper, we never observed the annihilation of vortices {\it via} the fold-Hopf mechanism with $\sigma=1$ (giving birth to two nodes), but always {\it via} the Bristol mechanism (fold-Hopf with $\sigma=-1$).}. The whole process, from the entrance until the exit of the nonlineary cell  ($0 \,\, {\rm mm}\,
\le z\le 70$ mm, or equivalently $0\le z/z_{\rm nl}\le 50$), is displayed in the attached video for the case $\varPhi_2=\pi$.

\section{Potential flow}

In this section we study the possibility to cast the fold-Hopf bifurcation under a potential form able to account for the presence of vortices. We show that this only possible if reducing the generic form \eqref{eq:foldhopf} to $\vec{v}=\vec{\nabla}S_{\rm f\hspace{0.1mm}H}$ where the expression of the potential flow $S_{\rm f\hspace{0.1mm}H}$ is given in \eqref{eq:Spaper}.

Let's consider a non-gradient dynamical system
\begin{equation}\label{eq:dyn1}
\dot{x}=f(x,y),\quad \dot{y}=g(x,y),\quad \mbox{with}\quad \partial_y f \neq \partial_x g,
\end{equation}
such as given by the normal form of the fold-Hopf bifurcation \cite{Guckenheimer2002,Kuznetsov2004}:
\begin{equation}
\label{eq:foldhopf}
    f(x,y) = \nu x+\alpha xy,\quad g(x,y) = \mu  + \sigma x^2 -y^2,
\end{equation}
where $\alpha$, $\nu$, $\mu \in \mathbb{R}$ and $\sigma =\pm 1$.  The system \eqref{eq:dyn1},\eqref{eq:foldhopf}
has at most four equilibria of coordinates $(0,\pm\sqrt{\mu })$ and 
\begin{equation}\label{eq:equil}
\left(\pm\sqrt{\sigma(\nu^2/\alpha^2-\mu )},-\nu/\alpha\right).
\end{equation}
The nature of the equilibrium points \eqref{eq:equil} depends -- when they exist --  on the relative sign of $\alpha$ and $\sigma$: if ${\rm sign}(\alpha)=\sigma$ these equilibria  are saddles, whereas if ${\rm sign}(\alpha)=-\sigma$ they are either centers (when $\nu=0$) or points surrounded by spirals, commonly termed {\it spirals}  (when $\nu\neq 0$) \cite{Guckenheimer2002}.

One can find a gradient dynamical system
\begin{equation}\label{eq:dyn2}
\dot{x}=\partial_x S,\quad \dot{y}=\partial_y S,
\end{equation}
 orbitally equivalent to \eqref{eq:dyn1}
if $S$ solves the linear first order partial derivative equation
\begin{equation}
\label{eq:S}
\frac{\partial_x S}{f(x,y)} = \frac{\partial_y S}{g(x,y)} \equiv h(x,y)>0 .
\end{equation}
The positivity of the integrating factor $h$ ensures that the direction of the flow described by the gradient system is identical to that of the original system \eqref{eq:dyn1}. Strictly speaking, $h(x,y)$ should also be a smooth function. However, in order to be able to describe flows containing vortices we authorize $h(x,y)$ to be singular. Eq.
\eqref{eq:S} can be solved by the method of characteristics, i.e., by writing $S(x,y) = {\cal S}(\xi)$ where $\xi(x,y)$ is constant along the characteristics, governed by the following ordinary differential equation
\begin{equation}\label{eq:char}
    \frac{dy}{dx} + \frac{f(x,y)}{g(x,y)}=0.
\end{equation}
We first note that the normal form \eqref{eq:foldhopf} is already a gradient system if $\alpha=2\sigma$. In this case $h=1$ and the velocity potential $S=\tfrac12 \nu x^2 + \mu  y +\sigma x^2 y -\tfrac13 y^3$ is continuously differentiable. It can be shown that such a flow does not admit closed orbits
\cite{strogatz_nonlinear_2015}: this case is of no interest for us since we want to describe vortices.

In the generic situation ($\alpha\neq 2\sigma$ and $h\neq 1$) the cases $\alpha=-\sigma$ and $\alpha=-2\sigma$
have to be treated separately.

\subsubsection{Standard situation}

Let's first consider the standard situation $\alpha\neq -2\sigma,-\sigma,2\sigma$.
Solving \eqref{eq:char} for the normal form of the fold-Hopf bifurcation~\eqref{eq:foldhopf} we obtain
\begin{equation*}
\begin{split}
    \xi = \left|\alpha y+\nu\right|^{2 \sigma/\alpha}
    \Big[\mu \left(\alpha^2\sigma+2\sigma+3 \alpha\right) -\sigma \mu^2\\
    +x^2 \left(\alpha^2+3 \alpha \sigma+2\right)-y^2 (\alpha+2 \sigma)+2 \nu y\Big].
\end{split}
\end{equation*}
yielding  the integrating factor
\begin{equation*}
    h = {\rm sgn}\,(\alpha y+\nu) {\cal S}'(\xi) (2\alpha^2+6\alpha \sigma+4)
    \left|\alpha y+\nu\right|^{\frac{2\sigma}{\alpha}-1} .
\end{equation*}
Since $\xi \propto  \left|\alpha y+\nu\right|^{2 \sigma/\alpha}$, ${\cal S}'(\xi)$ does not change sign on the line $y=\nu/\alpha$, such that $h$ necessarily changes sign at $y = \nu/\alpha$. Thus the inequality \eqref{eq:S} is violated and no gradient system orbitally equivalent to~\eqref{eq:dyn1},\eqref{eq:foldhopf} exists.

\subsubsection{The case $\alpha=-\sigma$}

If $\alpha=-\sigma$, one obtains
\begin{equation*}
    \xi = \ln \left|\sigma y-\nu\right|+\frac{\sigma x^2-2y^2+ \nu^2+\mu}{2 \left(\sigma y-\nu\right)^2},
\end{equation*}
yielding 
\begin{equation*}
    h =  \frac{-\sigma {\cal S}'(\xi)}{(\sigma y-\nu)^3}.
\end{equation*}
Similarly to the previous case, since $\xi \propto 1/(\sigma y-\nu)^2$, ${\cal S}'(\xi)$ does not change sign on the line $y=\sigma \nu$, and the integrating factor necessarily changes sign at $y = \sigma \nu$. Here also no gradient system orbitally equivalent to~\eqref{eq:dyn1},\eqref{eq:foldhopf} exists.

\subsubsection{The case $\alpha=-2 \sigma$}

If $\alpha=-2\sigma$, one obtains
\begin{equation}\label{eq:xi}
    \xi = \nu \ln \left|2 \sigma y-\nu\right|+2\,\frac{\sigma x^2+y^2+\mu-\nu^2/4}{2 \sigma y-\nu},
\end{equation}
yielding
\begin{equation*}
    h = \frac{-4\sigma {\cal S}'(\xi)}{(2 \sigma y-\nu)^2},
\end{equation*}
which is strictly positive provided that ${\rm sgn}[{\cal S}'(\xi)] = -\sigma$. Thus by picking an adequate function ${\cal S}$, we obtain a gradient system orbitally equivalent to~\eqref{eq:dyn1},\eqref{eq:foldhopf}. Notice that the points $(x,y)=(0,\pm \sqrt{\mu })$ remain equilibria, whereas points of coordinates \eqref{eq:equil} are now singularities of
$\vec{\nabla} S$.

Requiring that the circulation of $\vec{\nabla} S$ on a path encircling the singularities \eqref{eq:equil} is $\pm 2\pi$, we obtain
\begin{equation}
\label{eq:Sfh}
    S(x,y) = \arg[(2 y -\sigma\nu) (\xi+i \sigma)]\; .
\end{equation}
Note that the phase~\eqref{eq:Sfh} can be written as a piecewise function of $\tan^{-1}(\sigma/\xi)$ or equivalently $\cot^{-1}(\xi/\sigma)$.

For non zero values of $\nu$ the trajectories in the vicinity of the singularities are spirals. In this work we aim at describing vortices which are singularities of the flow surrounded by closed orbits. This imposes $\nu=0$. 

A word of caution is in order: It is well known that gradient systems do not admit closed orbits if the velocity potential is continuously differentiable
\cite{strogatz_nonlinear_2015}. This theorem does not apply here since the phase \eqref{eq:Sfh} has essential singularities and has a by construction $2\pi$-discontinuity
along the corresponding branch cut(s). 
This is the reason why it is possible to observed closed orbits around vortices in a gradient flow with velocity potential \eqref{eq:Sfh}.

In summary, in a quantum fluid, some mechanisms of vortices formation (or anihilation) can be qualitatively described by the function 
$S$ whose expression is given in \eqref{eq:Sfh} where $\xi$ is given by \eqref{eq:xi}. The resulting flow pattern is orbitally equivalent to the fold-Hopf bifurcation \eqref{eq:dyn1},\eqref{eq:foldhopf} relevant to vortex formation in the special case 
\begin{equation}\label{eq:fHok}
    \alpha=-2\sigma\quad\mbox{and}\quad \nu=0.
\end{equation}
In this case \eqref{eq:xi} reduces to
\begin{equation}\label{eq:xipaper}
\xi=\frac{\sigma x^2+y^2+\mu}{\sigma y},
\end{equation}
and the phase  \eqref{eq:Sfh} reads 
\begin{equation}\label{eq:Spaper}
S=S_{\rm f\hspace{0.1mm}H} \equiv\arg\big[x^2+\sigma(y^2+\mu)+ i \sigma y\big],
\end{equation}
which is the expression given in the main text. We remind that in this analogy, the variation of $\mu$ can be seen as the propagation along the $z$-axis. We will only consider this situation in the following.

\subsection{Bristol mechanism}

\begin{figure}
    \centering
\begin{picture}(8.5,4)
\put(0,0){\includegraphics[width=0.49\linewidth]{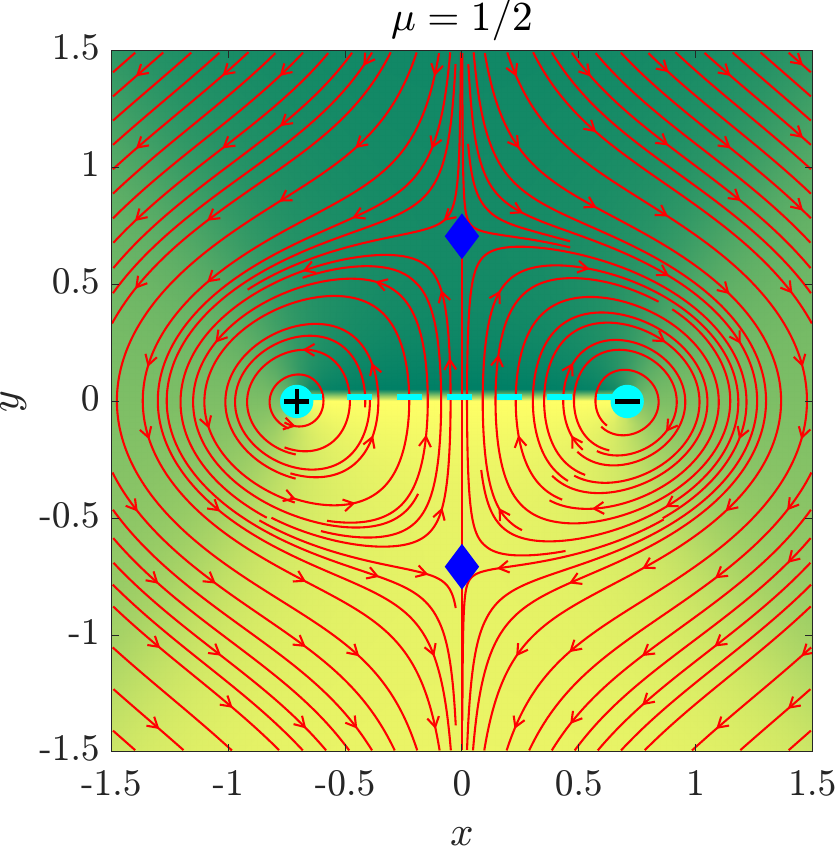}}
\put(4.4,0){\includegraphics[width=0.49\linewidth]{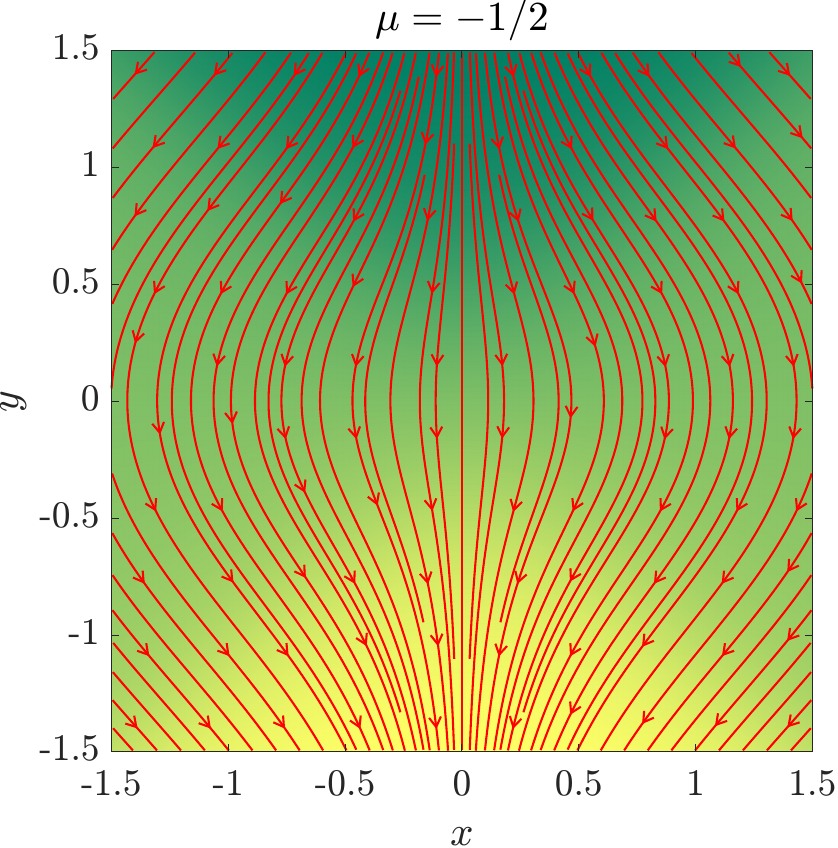}}
\put(0.7,4.2){(a)}
\put(5.1,4.2){(b)}
\end{picture}
\caption{Phase portraits of the dynamical system \eqref{eq:dyn1}, \eqref{eq:foldhopf}, \eqref{eq:fHok} [or equivalently \eqref{eq:dyn2},\eqref{eq:Spaper}] with $\sigma=-1$ for two different values of the bifurcation parameter $\mu$. Cyan circles are vortices, blue diamonds are saddles. The color scale corresponds here to the phase~\eqref{eq:Spaper} (light yellow corresponds to $S=\pi$ and dark green to $S=-\pi$). 
The position of the $2\pi$-jump of $S(x,y)$ [dashed line in panel (a)], is arbitrary and fixed by the choice of constant of integration in~\eqref{eq:S}.}\label{fig:S6}
\end{figure}

In this section we show that, in the case $\sigma=-1$, the system~\eqref{eq:dyn2},\eqref{eq:Spaper} describes the potential flow presented by the Bristol team in \cite{Nye1988} close to the bifurcation point ($x=y=\mu =0$). Indeed, the model equation (exact solution of the two-dimensional Helmholtz equation) obtained in~\cite{Nye1988} reads 
\begin{equation}
\label{eq:psib}
\psi_{\rm B}(x,y) =  (x^2-\mu -i y) \exp\{-i y\}.
\end{equation}
Let
\begin{equation}
    x = \sqrt{\varepsilon}\, X,\quad y = \sqrt{\varepsilon}\, Y,\quad \mu =\varepsilon M,
\end{equation}
such that the limit $\varepsilon \to 0$ corresponds to the bifurcation point. Expanding~\eqref{eq:psib} in $\varepsilon$ we obtain 
\begin{align}
    \psi_{\rm B}(x,y) &= -i \sqrt{\varepsilon}\, Y + \varepsilon(X^2-Y^2-M)+ O(\varepsilon^{3/2}),\nonumber\\
    &\sim x^2-y^2-\mu -iy.
\end{align}
Thus, in the neighbourhood of the bifurcation point, the phase of the solution~\eqref{eq:psib} is given by the potential~\eqref{eq:Spaper} derived previously with $\sigma=-1$.

The phase portrait of the gradient dynamical system~\eqref{eq:dyn2},\eqref{eq:Spaper} is displayed in Fig.~\ref{fig:S6} in the case $\sigma=-1$ which corresponds to the Bristol mechanism. The flow locally resembles the flow of the numerical simulation presented in Fig.~\ref{fig:S5} where two saddles annihilate with two vortices, yielding a flow without critical points for $\mu <0$.

\subsection{Incompressible flow}

In this section we make an attempt to describe the fold-Hopf bifurcation \eqref{eq:dyn1},\eqref{eq:foldhopf},\eqref{eq:fHok} for an incompressible flow. The discussion shows that the ``node to vortex bifurcation'' (i.e., fold-Hopf with $\sigma=1$) can be realized in an incompressible flow only at the expense of restrictions which make its experimental realization doubtful.

The condition of incompressibility is $\vec{\nabla}\cdot\vec{v}=0$, which, for a potential flow, implies that $S$ satisfies Laplace's equation. In our case $S(x,y)={\cal S}(\xi)$ where $\xi$ is given by  \eqref{eq:xipaper}, yielding
\begin{equation}\begin{split}
    \Delta S = & \frac{2}{y^2} [\xi+(1-\sigma)y] {\cal S}'(\xi)+\\
    & \frac{1}{y^2}[\xi^2-4\mu +4x^2(1-\sigma)] {\cal S}''(\xi)\; .
\end{split}
\end{equation}
Hence, when $\sigma=1$ (which is the case we consider henceforth) the condition of incompressibility $\Delta S=0$
can be expressed as an ordinary differential equation for the function ${\cal S}(\xi)$:
\begin{equation}\label{eq:odeH}
    2 \xi {\cal S}'(\xi)+ (\xi^2-4\mu ) {\cal S}''(\xi)=0\; .
\end{equation}
The function ${\cal S}$ solution of \eqref{eq:odeH} should verify
${\cal S}'(\xi)\propto (\xi^2-4\mu )^{-1}$ which leads to
\begin{equation}\label{eq:Hcases}
    {\cal S}(\xi)= \begin{cases}
\frac{1}{2} \ln 
\left|\frac{\displaystyle \xi-\sqrt{4\mu }}{\displaystyle \xi+\sqrt{4\mu }}\right|
& \mbox{if}\; \mu >0 \; , \vspace{3mm} \\
    \arctan (\xi/\sqrt{-4\,\mu }) & \mbox{if}\; \mu <0 \;  .
    \end{cases}
\end{equation}
The resulting velocity flow is
\begin{equation}\label{eq:vxy}\begin{split}
    v_x & =\frac{|4\mu |^{1/2}}{\xi^2-4\mu }\; \frac{2x}{y}\, \; ,\\ 
    v_y & =\frac{|4\mu |^{1/2}}{\xi^2-4\mu }
    \left(1-\frac{\mu +x^2}{y^2}\right)
\; ,
\end{split}
\end{equation}
and it is easy to check that, except at isolated exceptional points (see below)
\begin{equation}\label{divv0}
    \vec{\nabla}\cdot\vec{v}=0\; ,
\end{equation}
which is the condition of incompressibility for which function ${\cal S}$ \eqref{eq:Hcases} has been designed.

When $\mu <0$, the flow \eqref{eq:vxy} exhibits two vortices located at points of coordinates
$(\pm\sqrt{-\mu },0)$. 
At these points the velocity diverges and the phase is ill defined. One should check that the above solution verifies the Onsager-Feynman relation which implies that vortices are essential singularities of the phase connected by branch cuts along which $S$
experiences $2\pi$ jumps. Actually, this is pretty obvious from expression \eqref{eq:Hcases} in the case $\mu <0$, but it is instructive to perform the explicit computation.
Consider a circle of radius $\rho$ around, say, the vortex $(\sqrt{-\mu },0)$ with parametric equation $x=\sqrt{-\mu }+\rho\cos\theta$, $y=\rho\sin\theta$. For computing the circulation of the velocity field along this contour one needs to evaluate
\begin{equation}\label{eq:circul}
\begin{split}
&    \frac{1}{2\pi}\oint \vec{v}\cdot{\rm d}\vec{\ell}=\rho
    \int_0^{2\pi} \!\!(-v_x\sin\theta+v_y\cos\theta)\,  \frac{{\rm d}\theta}{2\pi}\\
&    =\frac{\sqrt{-\mu }}{\pi}\int_0^{2\pi}\!\!
\frac{-(2\sqrt{-\mu }+\rho\cos\theta){\rm d}\theta}{\rho^2-4\mu +4\rho\sqrt{-\mu }\cos\theta}=-1 \; .\end{split}
\end{equation}
This result holds only if $\rho<2\sqrt{-\mu }$. For larger values of $\rho$ the contour encloses the two vortices and the total circulation is zero as it should. Mathematically this stems from the formula
\begin{equation}
    \int_0^{2\pi}\frac{(1+e\cos\theta)\, {\rm d}\theta}{1+e^2+2\,e\cos\theta}=
    \begin{cases}
        2\pi & \mbox{if}\;\; |e|<1\; , \\
        \pi & \mbox{if}\;\; e=\pm 1\; , \\
        0 & \mbox{if}\;\; |e|>1\; ,
    \end{cases}
\end{equation}
where $e=\rho/(2\sqrt{-\mu })$ in \eqref{eq:circul}.

\begin{figure}
    \centering
    \includegraphics[width=\linewidth]{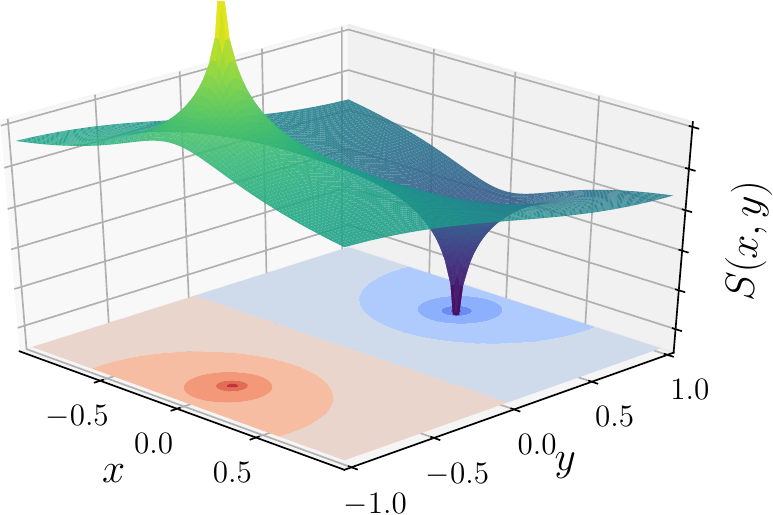}
    \caption{$S(x,y)={\cal S}(\xi)$ as given by \eqref{eq:Hcases} for $\mu =0.5$. The phase $S$ diverges at the nodes located at $\vec{r}_{\pm}=(0,\pm \sqrt{\mu })$. The lower horizontal plane displays a contour plot of $S$, where positive (negative) values are shown in red (blue).}
    \label{fig:S7}
\end{figure}

When $\mu >0$ the vortices disappear. From expression~\eqref{eq:vxy} one can check that there still exist two points $\vec{r}_{\pm}$, of coordinates $(0,\pm\sqrt{\mu })$, at which the velocity $|\vec{v}\,|$ diverges. These points are also extrema of the phase, with $S\to \pm\infty$. This divergence enables to reach a situation with extrema of $S$ at points where $\vec{\nabla} S\neq \vec{0}$, see Fig.~\ref{fig:S7}. Hence, apparently contrarily to what is stated in the main text, it seems (at least mathematically) possible to realize extrema of the phase within an incompressible fluid. However, this is obtained at the expense of the formation of points with a diverging velocity. The situation met in the main text is completely different from this idealized configuration: in the compressible quantum fluid we consider, phase extrema are indeed found to be stagnation points (stable or unstable nodes) at which the velocity cancels as expected.

To better understand the meaning of the extrema of $S$ in the context of an incompressible flow (when $\mu >0$), it is instructive to compute the integral of $\vec{\nabla}\cdot\vec{v}$ over a surface $\Omega_{\pm}$ enclosing the singular point $\vec{r}_{\pm}$. Since  $\vec{\nabla}\cdot\vec{v}=0$ except at this point, the shape of the curve delimiting the surface is immaterial (provided it does not also enclose the other singular point). For a circle of radius $\rho<2\sqrt{\mu }$, a simple use of the divergence theorem shows that 
\begin{equation}
    \int_{\Omega_{\pm}} \vec{\nabla}\cdot\vec{v} \; \dd x \dd y=\pm 2 \pi ,
\end{equation}
implying that, the flow \eqref{eq:vxy} verifies
\begin{equation}
    \vec{\nabla}\cdot\vec{v}=2\pi \delta(\vec{r}-\vec{r}_{+})
    -2\pi \delta(\vec{r}-\vec{r}_{-}),\;\;\mbox{when}\;\;
    \mu >0.
\end{equation}
This formula, which corrects and precises \eqref{divv0}, indicates that, in an incompressible fluid, it is possible to enforce the fold-Hopf bifurcation describing the birth of two nodes resulting from the annihilation of two vortices, only at the expense of the assumption that one of the nodes is a source and the other one a sink. Although quite appealing, such a configuration is certainly hardly realized in an  experimental setup.
This is the reason why we state in the main text that the new node to vortex bifurcation we identify cannot occur in an incompressible fluid.

\bibliography{biblio}